\documentclass[aps,prd,reprint,twocolumn,superscriptaddress]{revtex4-2}
\usepackage[english]{babel}
\usepackage[T1]{fontenc}
\usepackage{amsmath}
\usepackage{amssymb}
\usepackage{graphicx}
\usepackage{times}
\usepackage{euscript}
\usepackage{amsthm}
\usepackage{amsfonts}
\usepackage{multirow,makecell,array}
\usepackage{mathtools}
\usepackage{float}
\usepackage[dvipsnames]{xcolor}
\usepackage{bbm}

\begin{document}

\title{Vacuum polarization and pair production in time-dependent electric fields: \\ A quantum-kinetic-equation approach}

\author{I.~A.~Aleksandrov}
\email{i.aleksandrov@spbu.ru}
\affiliation{Department of Physics, Saint Petersburg State University, Universitetskaya Naberezhnaya 7/9, Saint Petersburg 199034, Russia}
\author{V.~A.~Bokhan}
\affiliation{Department of Physics, Saint Petersburg State University, Universitetskaya Naberezhnaya 7/9, Saint Petersburg 199034, Russia}
\author{A.~I.~Baksheev}
\affiliation{Department of Physics, Saint Petersburg State University, Universitetskaya Naberezhnaya 7/9, Saint Petersburg 199034, Russia}
\author{A.~Kudlis}
\affiliation{Science Institute, University of Iceland, Dunhagi 3, IS-107, Reykjavik, Iceland}

\begin{abstract}
The evolution of the vacuum state in a time-dependent external electric field of arbitrary polarization is investigated within a nonperturbative framework of quantum kinetic equations (QKEs). In our previous work [Phys. Rev. Res. {\bf 6}, 043009 (2024)], a revised version of the QKEs was derived by using an adiabatic basis constructed from one-particle Hamiltonian eigenfunctions in a spatially homogeneous electric field. In this study, we present an extensive analysis of these equations with particular emphasis on observable quantities. Specifically, we compute momentum-resolved particle yields, the induced electron-positron current, the energy-momentum tensor, and the angular-momentum tensor. We also discuss in detail the charge-renormalization procedure required to remove logarithmic divergences. It is shown that our results are consistent with the previous findings obtained via the Dirac-Heisenberg-Wigner formalism. Our analysis provides a firmer theoretical basis for investigations of nonperturbative effects in strong electric fields.
\end{abstract}

\maketitle

%%%%%%%%%%%%%%%%%%%%%%%%%%%%%%%%%%%%%%%%%%%%%%%%%%%%%%%%%

\section{Introduction}
\label{sec:intro}

A defining feature of quantum field theory---as opposed to ordinary quantum mechanics---is that particle number is not, in general, conserved in elementary processes. In quantum electrodynamics (QED), this manifests itself already in vacuum through fluctuations of virtual electron-positron pairs. It has long been predicted that, in a sufficiently strong constant electric field, these virtual excitations can be promoted to real $e^+e^-$ pairs~\cite{sauter_1931,heisenberg_euler,weisskopf,schwinger_1951}. This process, commonly referred to as the Sauter-Schwinger mechanism, is intrinsically nonperturbative in the field strength, which makes its experimental observation one of the central goals in the context of strong-field QED (see, e.g., reviews~\cite{dipiazza_rmp_2012,xie_review_2017,gonoskov_2022,fedotov_review}).

Since realistic scenarios involve external fields with nontrivial space-time structure, the theoretical description of the vacuum decay with production of real particles requires robust nonperturbative techniques. Modern numerical methods that exactly incorporate the interaction with the external background fields are mainly based on several general frameworks. First, the quantitative analysis of the pair-production process can be carried out by solving the Dirac equation in the presence of the external field. The particle spectra can be extracted according to the theory with unstable vacuum which relies on the Furry-picture quantization~\cite{fradkin_gitman_shvartsman,gavrilov_prd_1996}. This approach has been implemented in various forms in many works (see, e.g., Refs.~\cite{gavrilov_prd_1996,ruf_prl_2009,woellert_prd_2015,aleksandrov_prd_2016,aleksandrov_prd_2017_1,aleksandrov_prd_2017,lv_pra_2018,lv_prl_2018,aleksandrov_prd_2018,maltsev_prl_2019,peng_prr_2020,aleksandrov_kohlfuerst,sevostyanov_prd_2021,majczak_prd_2024,su_prd_2025}). An alternative is to evolve the covariant Wigner function using the Dirac-Heisenberg-Wigner (DHW) equations~\cite{vasak_1987,BB_prd_1991,zhuang_1996,zhuang_prd_1998,ochs_1998,hebenstreit_prd_2010,fauth_prd_2021} (see, e.g., Refs.~\cite{aleksandrov_kohlfuerst,hebenstreit_prl_2009,blinne_gies_2014,blinne_strobel_2016,olugh_prd_2019,kohlfuerst_prd_2019,li_prd_2019,kohlfuerst_prd_2020,kohlfuers_prdl_2024,kohlfuerst_arxiv_2022,yu_prd_2023,hu_prd_2023,hu_prd_2024,majczak_prd_2024,aleksandrov_kudlis_klochai,chen_prd_2025,bake_prd_2025,jiang_arxiv_2025}). In many laser-driven setups, the external field can be approximated as spatially homogeneous, i.e., depending only on time in a given frame (see a recent study~\cite{tkachev_pra_2025} and references therein). In this case, a computationally efficient description of the pair-production phenomenon is provided by quantum kinetic equations (QKEs)~\cite{mamaev_trunov_1979,mostepanenko_1980,GMM,pervushin_2005,pervushin_skokov,schmidt_1998,kluger_prd_1998,schmidt_prd_1999,bloch_prd_1999,dumlu_prd_2009,fedotov_prd_2011,blaschke_prd_2011,aleksandrov_epjst_2020,aleksandrov_kudlis_klochai} (see also Refs.~\cite{sevostyanov_prd_2021,alkofer_prl_2001,blaschke_prl_2006,otto_plb_2015,panferov_epjd_2016,aleksandrov_symmetry,aleksandrov_sevostyanov_2025,aleksandrov_kudlis_prdl_2024,edwards_prdl_2025,li_arxiv_2025,brass_arxiv_2025}).

In our recent work~\cite{aleksandrov_kudlis_klochai}, we revisited the QKE formulation for time-dependent electric fields with \emph{arbitrary} polarization. We showed that a self-consistent kinetic system differs from the form previously used in the literature~\cite{pervushin_skokov} and, in particular, can be written as a closed set of ten ordinary differential equations. Moreover, we established explicit linear relations between the QKE components and the DHW functions, thereby demonstrating the equivalence of the two descriptions in the spatially homogeneous case~\cite{aleksandrov_kudlis_klochai}.

The present paper builds on these results and provides a detailed analysis of the QKE framework and the associated physical observables. We focus on momentum-resolved particle spectra and on vacuum-induced expectation values such as the electron-positron current, the energy-momentum tensor, and the angular-momentum tensor. A central aspect of this program is the proper treatment of ultraviolet behavior: several observables exhibit logarithmic divergences that must be removed by charge renormalization. We therefore scrutinize the renormalization procedure in detail and formulate practical subtraction prescriptions while preserving gauge invariance. In the case of a linearly polarized external field, ultraviolet divergences were analyzed and removed in Refs.~\cite{mamaev_trunov_1979,mostepanenko_1980,GMM} using a subtraction scheme originally introduced in the context of cosmological pair production~\cite{zeldovich_1972}. The corresponding renormalized current density was subsequently studied in a number of specific linearly polarized setups; see, e.g., Refs.~\cite{bloch_prd_1999,aleksandrov_pra_2021}. While the final subtraction prescriptions are stated in the aforementioned works, the intermediate derivation steps are not spelled out in detail. We therefore (i)~extend the analysis to the more general case of arbitrary polarization and (ii)~provide a self-contained derivation of the subtraction procedure based on Pauli-Villars regularization~\cite{pauli_villars_1949}. The resulting renormalized quantities are shown to be consistent with the known expressions obtained within the DHW formalism, thereby strengthening the theoretical basis for quantitative studies of nonperturbative QED effects in strong, time-dependent electric fields.

This paper is organized as follows. In Sec.~\ref{sec:derivation} we outline the derivation of the QKE system for an arbitrarily polarized, spatially homogeneous electric background, emphasizing the role of the adiabatic basis and the evolution equations for the relevant correlation functions. In Sec.~\ref{sec:PT} we analyze the weak-field (perturbative) regime and derive leading-order expressions used later to identify a subtraction term for the current density. In Sec.~\ref{sec:unrenorm} we express the unrenormalized current, energy-momentum tensor, and angular-momentum tensor in terms of the QKE functions, discuss their physical interpretation, and examine the ultraviolet asymptotics. Section~\ref{sec:renorm} is devoted to renormalization, where we present and compare two complementary approaches and obtain finite, gauge-invariant expressions for the renormalized observables. In Sec.~\ref{sec:LP} we summarize the simplified form of the QKEs and the observables for the special case of linear polarization. We conclude in Sec.~\ref{sec:conclusion}. Throughout the paper we use units $\hbar=c=1$, $\alpha=e^2/(4\pi)$, and take $e<0$. The metric convention is $g^{\mu \nu} = \mathrm{diag} (1, -1, -1, -1)$.

%%%%%%%%%%%%%%%%%%%%%%%%%%%%%%%%%%%%%%%%%%%%%%%%%%%%%%%%%

\section{Derivation of the QKE system} \label{sec:derivation}

\subsection{General framework and notation}

Here we briefly outline the main steps in deriving the QKE system. Our approach differs slightly from that of Ref.~\cite{aleksandrov_kudlis_klochai}. Although it will also revolve around the concept of the adiabatic basis, instead of the Dirac equation for one-particle solutions, we will invoke the Heisenberg equations for the creation and annihilation operators. The latter will allow us to directly obtain the equations governing the temporal evolution of the relevant correlation functions.

We describe the external electric field $\mathbf{E}(t)$ in the temporal gauge $A_0 = 0$, $\mathbf{E}(t) = - \dot{\mathbf{A}}(t)$. We assume that the electric field vanishes for $t\leqslant t_\text{in}$ and $t \geqslant t_\text{out}$. We choose the vector potential such that $\mathbf{A} (t\leqslant t_\text{in}) = \mathbf{A}_\text{in}$ and $\mathbf{A} (t\geqslant t_\text{out}) = 0$. Accordingly, the one-particle Hamiltonian in the coordinate representation reads $\mathcal{H}_e (t) = \boldsymbol{\alpha} [-i \boldsymbol{\nabla} - e \mathbf{A} (t)] + \beta m$. For each fixed $t$, this Hamiltonian is unitarily equivalent to the free Dirac Hamiltonian. We introduce the adiabatic eigenfunctions:
\begin{equation}
\mathcal{H}_e (t) \varphi^{(\zeta)}_{\mathbf{p},s} (\mathbf{x}; t) = \zeta \omega(\zeta \mathbf{p}, t) \varphi^{(\zeta)}_{\mathbf{p},s} (\mathbf{x}; t), \quad \zeta = \pm,
\label{eq:adiabatic_eigenfuncions}
\end{equation}
where
\begin{equation}
\omega (\mathbf{p}, t) = \sqrt{m^2 + [\mathbf{p} - e \mathbf{A}(t)]^2} \label{eq:omega_def}    
\end{equation}
and $s$ labels the spin state. In Eq.~\eqref{eq:adiabatic_eigenfuncions} the semicolon indicates that the $t$ dependence is parametric. In the explicit form, the functions of the adiabatic basis read
\begin{align}
\varphi^{(+)}_{\mathbf{p},s} (\mathbf{x}; t) &= (2\pi)^{-3/2} \, \mathrm{e}^{i\mathbf{p} \mathbf{x}} \, u_{\mathbf{p}-e\mathbf{A} (t),s}, \label{eq:adiabatic_plus}\\
\varphi^{(-)}_{\mathbf{p},s} (\mathbf{x}; t) &= (2\pi)^{-3/2} \, \mathrm{e}^{-i\mathbf{p} \mathbf{x}} \, v_{-\mathbf{p}-e\mathbf{A} (t),s},\label{eq:adiabatic_minus}
\end{align}
where the bispinors $u_{\mathbf{p},s}$ and $v_{\mathbf{p},s}$ will be specified below. Since the Heisenberg field operator $\psi (x)$ obeys an equation of motion of the same form as the Dirac equation in a given external field, it is convenient to expand $\psi (x)$ in the adiabatic basis:
\begin{equation}
\psi (x) = \sum_{s} \int \! d\mathbf{p} \big [a_{\mathbf{p},s} (t) \varphi^{(+)}_{\mathbf{p},s} (\mathbf{x}; t) + b^\dagger_{\mathbf{p},s} (t) \varphi^{(-)}_{\mathbf{p},s} (\mathbf{x}; t) \big ]. \label{eq:psi_adiabatic}
\end{equation}
The associated creation and annihilation operators are time-dependent because the adiabatic functions are not solutions of the time-dependent Dirac equation. However, at any given $t$, the adiabatic operators obey the usual anticommutation relations. We denote the exact Dirac solutions by ${}_\zeta \varphi_{\mathbf{p},s} (x)$ and ${}^\zeta \varphi_{\mathbf{p},s} (x)$. These two sets differ because the {\it in} solutions ${}_\zeta \varphi_{\mathbf{p},s} (x)$ are fixed by their asymptotic form for $t \leqslant t_\text{in}$, while the {\it out} solutions are fixed by their asymptotic behavior for $t \geqslant t_\text{out}$. Namely,
\begin{align}
\mathcal{H}_e (t_\text{in}) {}_\zeta \varphi_{\mathbf{p},s} (t_\text{in}, \mathbf{x}) &= \zeta \omega (\zeta \mathbf{p}, t_\text{in}) {}_\zeta \varphi_{\mathbf{p},s} (t_\text{in}, \mathbf{x}), \\
\mathcal{H}_e (t_\text{out}) {}^\zeta \varphi_{\mathbf{p},s} (t_\text{out}, \mathbf{x}) &= \zeta \omega (\zeta \mathbf{p}, t_\text{out}) {}^\zeta \varphi_{\mathbf{p},s} (t_\text{out}, \mathbf{x}),
\end{align}
where, in the gauge chosen above, $\omega (\mathbf{p}, t_\text{in}) = \sqrt{m^2 + (\mathbf{p} - e \mathbf{A}_\text{in})^2}$ and $\omega (\mathbf{p}, t_\text{out}) = \sqrt{m^2 + \mathbf{p}^2} \equiv p^0 = p_0$.

When $\psi(x)$ is expanded in in/out solutions, the corresponding creation and annihilation operators are time independent. We denote them by $a_{\mathbf{p},s} (\text{in})$, $b^\dagger_{\mathbf{p},s} (\text{in})$ and $a_{\mathbf{p},s} (\text{out})$, $b^\dagger_{\mathbf{p},s} (\text{out})$, respectively. According to our definitions,
\begin{equation*}
\varphi^{(\zeta)}_{\mathbf{p},s} (\mathbf{x}; t_\text{in}) = {}_\zeta \varphi_{\mathbf{p},s} (t_\text{in}, \mathbf{x}), \quad \varphi^{(\zeta)}_{\mathbf{p},s} (\mathbf{x}; t_\text{out}) = {}^\zeta \varphi_{\mathbf{p},s} (t_\text{out}, \mathbf{x}).
\end{equation*}
Thus, at asymptotic times the adiabatic functions coincide with the corresponding in/out solutions. It then follows that
\begin{align*}
a_{\mathbf{p},s} (t_\text{in}) &= a_{\mathbf{p},s} (\text{in}) \,, \quad a_{\mathbf{p},s} (t_\text{out}) = a_{\mathbf{p},s} (\text{out}) \,, \\
b^\dagger_{\mathbf{p},s} (t_\text{in}) &= b^\dagger_{\mathbf{p},s} (\text{in}) \,, \quad b^\dagger_{\mathbf{p},s} (t_\text{out}) = b^\dagger_{\mathbf{p},s} (\text{out}) \,.
\end{align*}
The final electron distributions can be found via
\begin{equation}
n_{\mathbf{p},s}^{(e^-)} = \langle 0, \text{in} | a^\dagger_{\mathbf{p},s} (\text{out}) a_{\mathbf{p},s} (\text{out}) | 0, \text{in} \rangle,
\label{eq:n_gen}
\end{equation}
where $| 0, \text{in} \rangle$ is the {\it in} vacuum state: $a_{\mathbf{p},s} (\text{in})| 0, \text{in} \rangle = b_{\mathbf{p},s} (\text{in})| 0, \text{in} \rangle = 0$. Our derivation of the QKE system will be based on the following observation. If one defines the correlation functions of the type~\eqref{eq:n_gen} in terms of the adiabatic operators, then at $t = t_\text{out}$ they will yield the final momentum distributions of the particles produced by the external field. Our goal is to obtain a closed system governing the time evolution of the adiabatic correlation functions.

\subsection{Adiabatic correlation functions}

The in operators are related to the adiabatic ones by a time-dependent unitary (Bogoliubov) transformation. Its coefficients also encode the relation between the in solutions and the adiabatic basis. Since the external field does not depend on the spatial coordinates, these coefficients are (anti)diagonal with respect to the generalized momentum $\mathbf{p}$. Although one can obtain their time dependence directly from the Dirac equation~\cite{aleksandrov_kudlis_klochai}, we will follow a different route.

We consider the following expectation values with respect to the in vacuum:
\begin{widetext}
\begin{align}
    \langle 0, \text{in}|a^{\dagger}_{\mathbf{p},s}(t)a_{\mathbf{p}',s'}(t)|0, \text{in}\rangle &=
    \delta(\mathbf{p} - \mathbf{p}')A_{s s'}(\mathbf{p},t),
\label{eq:_A} \\
    \langle 0, \text{in}|b^{\dagger}_{-\mathbf{p},s}(t)b_{-\mathbf{p'},s'}(t)|0, \text{in}\rangle &=
    \delta(\mathbf{p} - \mathbf{p}')B_{s s'}(\mathbf{p},t),
\label{eq:_B} \\
    \langle 0, \text{in}|a^{\dagger}_{\mathbf{p},s}(t)b^{\dagger}_{-\mathbf{p'},s'}(t)|0, \text{in}\rangle &= \delta(\mathbf{p} - \mathbf{p}')C_{s s'}(\mathbf{p},t),
\label{eq:_C}\\
    \langle 0, \text{in}|b_{-\mathbf{p},s}(t)a_{\mathbf{p}',s'}(t)|0, \text{in}\rangle &=
    \delta(\mathbf{p} - \mathbf{p}')D_{s s'}(\mathbf{p},t).
\label{eq:_D}
\end{align}
\end{widetext}
The delta-functions appear due to momentum conservation mentioned above. According to Eqs.~\eqref{eq:n_gen} and \eqref{eq:_A}, the final electron number density is given by
\begin{equation}
n_{\mathbf{p},s}^{(e^-)} = \frac{V}{(2\pi)^3} \, A_{ss} (\mathbf{p}, t_\text{out}), \label{eq:np_A}
\end{equation}
where $V$ is the normalization volume (it is the number of particles per unit volume that yields a finite value). In order to obtain the system of equations governing the functions $A$, $B$, $C$, and $D$, we will first construct the differential equations involving the adiabatic creation and annihilation operators themselves. To this end, let us consider the following inner product:
\begin{align}
a_{\mathbf{p},s}(t) &= (\varphi^{(+)}_{\mathbf{p},s}(\mathbf{x};t), \psi (x)) \nonumber \\
{}&\equiv \int \! d\mathbf{x} \, \big [\varphi^{(+)}_{\mathbf{p},s} (\mathbf{x};t) \big ]^{\dagger} \psi (x).
\label{a:inner_product}
\end{align}
This relation holds due to the fact that the adiabatic functions form an orthonormal set of functions. Now we can directly calculate the time derivative of Eq.~\eqref{a:inner_product}. The field operator obeys
\begin{equation}
\dot{\psi} (x) = -i [ \psi(x), H(t)] = -i \mathcal{H}_e (t) \psi (x).
\end{equation}
Moreover, the adiabatic functions~\eqref{eq:adiabatic_plus} and \eqref{eq:adiabatic_minus} are known explicitly, so their derivatives are also known. Let us specify the bispinors. First, we note that they must obey
\begin{align}
\big ( \boldsymbol{\alpha} \mathbf{p} + \beta m \big ) u_{\mathbf{p},s} &= p^0 u_{\mathbf{p},s} \,, \label{eq:bispinors_eqs_u} \\
\big ( \boldsymbol{\alpha} \mathbf{p} + \beta m \big ) v_{\mathbf{p},s} &= -p^0 v_{\mathbf{p},s}. \label{eq:bispinors_eqs_v}
\end{align}
The bispinors $u_{\mathbf{p},s}$ correspond to the positive-energy eigenfunctions~\eqref{eq:adiabatic_plus} and they are fixed so far only up to an arbitrary unitary transformation. However, it turns out that it is very convenient to choose them in the following form (see Ref.~\cite{aleksandrov_kudlis_klochai}, where the ambiguity of the bispinor basis is discussed in more detail):
\begin{align}
u_{\mathbf{p},-1} &= C(p^0)
\begin{pmatrix}
p^0 + m\\
0\\
p_z\\
p_x+ i p_y
\end{pmatrix}, \label{eq:u_explicit_1} \\
u_{\mathbf{p},+1} &= C(p^0)
\begin{pmatrix}
0\\
p^0 + m\\
p_x - i p_y\\
-p_z
\end{pmatrix},\label{eq:u_explicit_2}
\end{align}
where $C(p^0) = [2p^0 (p^0 +m)]^{-1/2}$. The negative-energy eigenfunctions~\eqref{eq:adiabatic_minus} contain $v_{\mathbf{p},s}$ specified according to
\begin{align}
v_{\mathbf{p},-1} &= C(p^0)
\begin{pmatrix}
-p_z\\
-p_x - i p_y\\
p^0 + m\\
0
\end{pmatrix}, \label{eq:v_explicit_1} \\
v_{\mathbf{p},+1} &= C(p^0)
\begin{pmatrix}
-p_x + i p_y\\
p_z\\
0\\
p^0 + m
\end{pmatrix}.\label{eq:v_explicit_2}
\end{align}
The bispinors possess the following properties:
\begin{align}
&u^\dagger_{\mathbf{p},s} u_{\mathbf{p},s'} = v^\dagger_{\mathbf{p},s} v_{\mathbf{p},s'} = \delta_{ss'} \,, \quad u^\dagger_{\mathbf{p},s} v_{\mathbf{p},s'} = 0 , \\
&\sum_{s=\pm 1} \big ( u_{\mathbf{p},s} u^\dagger_{\mathbf{p},s} + v_{\mathbf{p},s} v^\dagger_{\mathbf{p},s}\big ) = 1, \\
&\sum_{s=\pm 1}  u_{\mathbf{p},s} \overline{u}_{\mathbf{p},s} = \frac{\gamma^0 p^0  - \boldsymbol{\gamma} \mathbf{p} + m}{2p^0} , \\
&\sum_{s=\pm 1} v_{\mathbf{p},s} \overline{v}_{\mathbf{p},s} = \frac{\gamma^0 p^0  + \boldsymbol{\gamma} \mathbf{p} - m}{2p^0},
\end{align}
where $\overline{u} \equiv u^\dagger \gamma^0$. Now one can straightforwardly demonstrate
\begin{widetext}
\begin{equation}
\dot{a}_{\mathbf{p},s}(t) = -i \omega (\mathbf{p},t) a_{\mathbf{p},s}(t) - \sum_{s'} \left[ i \boldsymbol{\mu}_1 (\mathbf{p},t) \boldsymbol{\sigma}_{s s'} a_{\mathbf{p},s'}(t) + \boldsymbol{\mu}_2 (\mathbf{p},t) \boldsymbol{\sigma}_{s s'} b^{\dagger}_{-\mathbf{p},s'}(t) \right],
\label{eq:a_dot}
\end{equation}
where $\boldsymbol{\sigma}$ is a vector containing the Pauli matrices, whose elements are indexed by $s$, $s' = -1$, $+1$. Besides,
\begin{align}
\boldsymbol{\mu}_1 (\mathbf{p},t) &= \frac{e}{2\omega (\omega+m)}\, [\mathbf{q} \times \mathbf{E} (t)], \label{eq:mu1} \\
\boldsymbol{\mu}_2 (\mathbf{p},t) &= \frac{e}{2\omega^2 (\omega + m)}\, \big \{ [\mathbf{q} \mathbf{E}(t)] \mathbf{q} - \omega (\omega + m) \mathbf{E} (t) \big \}, \label{eq:mu2} 
\end{align}
where $\mathbf{q} = \mathbf{q} (t) = \mathbf{p} - e\mathbf{A}(t)$ is the kinetic momentum. Along the same lines, one can obtain
\begin{equation}
\dot{b}^\dagger_{-\mathbf{p},s}(t) = i \omega (\mathbf{p},t) b^\dagger_{-\mathbf{p},s}(t) + \sum_{s'} \left[ \boldsymbol{\mu}_2 (\mathbf{p},t) \boldsymbol{\sigma}_{s s'} a_{\mathbf{p},s'}(t) - i \boldsymbol{\mu}_1 (\mathbf{p},t) \boldsymbol{\sigma}_{s s'} b^{\dagger}_{-\mathbf{p},s'}(t) \right].
\label{eq:b_dot}
\end{equation}
In what follows, we will explicitly differentiate Eqs.~\eqref{eq:_A}--\eqref{eq:_D} and employ the relations~\eqref{eq:a_dot} and \eqref{eq:b_dot}.

\subsection{Quantum kinetic equations}

By differentiating Eq.~\eqref{eq:_A}, we find
\begin{align}
\dot{A}_{s s'}(\mathbf{p},t) &= i \boldsymbol{\mu}_1 (\mathbf{p},t) \sum_{s''} \boldsymbol{\sigma}_{s'' s} A_{s'' s'} (\mathbf{p},t) - i \boldsymbol{\mu}_1 (\mathbf{p},t) \sum_{s''} \boldsymbol{\sigma}_{s' s''} A_{s s''} (\mathbf{p},t) \nonumber \\
{}&- \boldsymbol{\mu}_2 (\mathbf{p},t) \sum_{s''} \boldsymbol{\sigma}_{s'' s} D_{s'' s'} (\mathbf{p},t) - \boldsymbol{\mu}_2 (\mathbf{p},t) \sum_{s''} \boldsymbol{\sigma}_{s' s''} C_{s s''} (\mathbf{p},t). \label{eq:A_dot}
\end{align}
We omit the common factor $\delta (\mathbf{p} - \mathbf{p}')$ and set $\mathbf{p}' = \mathbf{p}$. Proceeding analogously, we obtain
\begin{align}
\dot{B}_{s s'}(\mathbf{p},t) &= i \boldsymbol{\mu}_1 (\mathbf{p},t) \sum_{s''} \boldsymbol{\sigma}_{s'' s'} B_{s s''} (\mathbf{p},t) - i \boldsymbol{\mu}_1 (\mathbf{p},t) \sum_{s''} \boldsymbol{\sigma}_{s s''} B_{s'' s'} (\mathbf{p},t) \nonumber \\
{}&- \boldsymbol{\mu}_2 (\mathbf{p},t) \sum_{s''} \boldsymbol{\sigma}_{s s''} D_{s' s''} (\mathbf{p},t) - \boldsymbol{\mu}_2 (\mathbf{p},t) \sum_{s''} \boldsymbol{\sigma}_{s'' s'} C_{s'' s} (\mathbf{p},t), \label{eq:B_dot} \\
\dot{C}_{s s'}(\mathbf{p},t) &= 2i\omega (\mathbf{p},t) C_{s s'}(\mathbf{p},t) + i \boldsymbol{\mu}_1 (\mathbf{p},t) \sum_{s''} \boldsymbol{\sigma}_{s'' s} C_{s'' s'} (\mathbf{p},t) - i \boldsymbol{\mu}_1 (\mathbf{p},t) \sum_{s''} \boldsymbol{\sigma}_{s' s''} C_{s s''} (\mathbf{p},t) \nonumber \\
{}&+ \boldsymbol{\mu}_2 (\mathbf{p},t) \sum_{s''} \boldsymbol{\sigma}_{s'' s} B_{s' s''} (\mathbf{p},t) + \boldsymbol{\mu}_2 (\mathbf{p},t) \sum_{s''} \boldsymbol{\sigma}_{s' s''} A_{s s''} (\mathbf{p},t) - \boldsymbol{\mu}_2 (\mathbf{p},t) \boldsymbol{\sigma}_{s' s}, \label{eq:C_dot} \\
\dot{D}_{s s'}(\mathbf{p},t) &= -2i\omega (\mathbf{p},t) D_{s s'}(\mathbf{p},t) + i \boldsymbol{\mu}_1 (\mathbf{p},t) \sum_{s''} \boldsymbol{\sigma}_{s'' s} D_{s'' s'} (\mathbf{p},t) - i \boldsymbol{\mu}_1 (\mathbf{p},t) \sum_{s''} \boldsymbol{\sigma}_{s' s''} D_{s s''} (\mathbf{p},t) \nonumber \\
{}&+ \boldsymbol{\mu}_2 (\mathbf{p},t) \sum_{s''} \boldsymbol{\sigma}_{s'' s} A_{s'' s'} (\mathbf{p},t) + \boldsymbol{\mu}_2 (\mathbf{p},t) \sum_{s''} \boldsymbol{\sigma}_{s' s''} B_{s'' s} (\mathbf{p},t) - \boldsymbol{\mu}_2 (\mathbf{p},t) \boldsymbol{\sigma}_{s' s}. \label{eq:D_dot}
\end{align}
Definitions~\eqref{eq:_A}--\eqref{eq:_D} imply the symmetry properties: $A^*_{s s'}(\mathbf{p},t) = A_{s' s}(\mathbf{p},t)$, $B^*_{s s'}(\mathbf{p},t) = B_{s' s}(\mathbf{p},t)$, and $C^*_{s s'}(\mathbf{p},t) = D_{s' s}(\mathbf{p},t)$. In other words, $A$ and $B$ are Hermitian, $A^\dagger = A$, whereas $C^\dagger = D$. We parameterize the correlation functions in the following form:
\begin{align}
A_{s s'}(\mathbf{p},t) &= a_0(\mathbf{p},t) \delta_{s s'} + \mathbf{a}(\mathbf{p},t)\boldsymbol{\sigma}_{s' s}, \label{eq:Asigma} \\
B_{s s'}(\mathbf{p},t) &= b_0(\mathbf{p},t) \delta_{s s'} + \mathbf{b}(\mathbf{p},t)\boldsymbol{\sigma}_{s s'}, \label{eq:Bsigma}\\
C_{s s'}(\mathbf{p},t) &= [u(\mathbf{p},t) - iv (\mathbf{p},t)] \delta_{s s'} + [ \mathbf{u} (\mathbf{p},t) - i \mathbf{v} (\mathbf{p},t) ] \boldsymbol{\sigma}_{s' s}, \label{eq:Csigma}\\
D_{s s'}(\mathbf{p},t) &= [u(\mathbf{p},t) + iv (\mathbf{p},t)] \delta_{s s'} + [ \mathbf{u} (\mathbf{p},t) + i \mathbf{v} (\mathbf{p},t) ] \boldsymbol{\sigma}_{s' s}. \label{eq:Dsigma}
\end{align}
Here the 16 unknown components are real and vanish at $t \leqslant t_\text{in}$. By substituting Eqs.~\eqref{eq:Asigma}--\eqref{eq:Dsigma} into Eqs.~\eqref{eq:A_dot}--\eqref{eq:D_dot} and using the identity
\begin{equation}
\sum_{s''} (\mathbf{a} \boldsymbol{\sigma}_{s s''})(\mathbf{b} \boldsymbol{\sigma}_{s'' s'}) = (\mathbf{a} \mathbf{b})\delta_{s s'} + i(\mathbf{a}\times\mathbf{b})\boldsymbol{\sigma}_{s s'},
\end{equation}
we obtain the following system of equations:
\begin{align}
\dot{a_0} &= -2 \boldsymbol{\mu}_2 \mathbf{u}, \label{eq:__a0}\\
\dot{b_0} &= -2 \boldsymbol{\mu}_2 \mathbf{u}, \label{eq:__b0}\\
\dot{\mathbf{a}} &= -2\boldsymbol{\mu}_2 u + 2(\boldsymbol{\mu}_1\times\mathbf{a}) - 2(\boldsymbol{\mu}_2\times\mathbf{v}), \label{eq:__a}\\
\dot{\mathbf{b}} &= -2 \boldsymbol{\mu}_2 u + 2(\boldsymbol{\mu}_1\times\mathbf{b}) + 2(\boldsymbol{\mu}_2\times\mathbf{v}), \label{eq:__b}\\
\dot{u}-i\dot{v} &= 2i\omega(u - iv)+\boldsymbol{\mu}_2 (\mathbf{a} + \mathbf{b}), \label{eq:__u+iv}\\
\dot{\mathbf{u}} - i\dot{\mathbf{v}} &= \boldsymbol{\mu}_2 (a_0 + b_0 - 1) + 2i\omega(\mathbf{u} - i\mathbf{v}) + 2(\boldsymbol{\mu}_1\times\mathbf{u})\nonumber \\
{}&- 2i(\boldsymbol{\mu}_1\times\mathbf{v}) + i(\boldsymbol{\mu}_2\times\mathbf{a})-i(\boldsymbol{\mu}_2\times\mathbf{b}). \label{eq:__vectr_u+iv}
\end{align}
The last two equations can be, of course, separately formulated for their real and imaginary parts, so there are 16 real equations. By taking into account the zero initial conditions and using Eqs.~\eqref{eq:__a0} and \eqref{eq:__b0}, we find $a_0 (\mathbf{p}, t) = b_0 (\mathbf{p}, t) \equiv f (\mathbf{p}, t)$. Furthermore, Eq.~\eqref{eq:__u+iv} together with a sum of Eqs.~\eqref{eq:__a} and \eqref{eq:__b} form a closed homogeneous subsystem for $u$, $v$, and $\mathbf{a} + \mathbf{b}$. This means that $u = v = 0$ and $\mathbf{a} (\mathbf{p}, t) = -\mathbf{b} (\mathbf{p}, t) \equiv \mathbf{f} (\mathbf{p}, t)$. We thus arrive at
\begin{align}
\dot{f} &= -2 \boldsymbol{\mu}_2 \mathbf{u}, \label{eq:qke_f}\\
\dot{\mathbf{f}} &= 2(\boldsymbol{\mu}_1\times\mathbf{f}) - 2(\boldsymbol{\mu}_2\times\mathbf{v}), \label{eq:qke_fvec}\\
\dot{\mathbf{u}} &= 2(\boldsymbol{\mu}_1\times\mathbf{u}) + \boldsymbol{\mu}_2 (2f - 1) + 2\omega \mathbf{v}, \label{eq:qke_uvec} \\
\dot{\mathbf{v}} &= 2(\boldsymbol{\mu}_1\times\mathbf{v}) - 2(\boldsymbol{\mu}_2\times\mathbf{f}) - 2 \omega \mathbf{u}. \label{eq:qke_vvec}
\end{align}
This reproduces the QKE system of Ref.~\cite{aleksandrov_kudlis_klochai}. It involves ten real functions that vanish at $t = t_\text{in}$. For later use, we summarize the correlation functions:
\begin{align}
    \langle 0, \text{in}|a^{\dagger}_{\mathbf{p},s}(t)a_{\mathbf{p}',s'}(t)|0, \text{in}\rangle &=
    \delta(\mathbf{p} - \mathbf{p}') [f(\mathbf{p},t) \delta_{s s'} + \mathbf{f}(\mathbf{p},t)\boldsymbol{\sigma}_{s' s}],
\label{eq:_A2} \\
    \langle 0, \text{in}|b^{\dagger}_{-\mathbf{p},s}(t)b_{-\mathbf{p'},s'}(t)|0, \text{in}\rangle &=
    \delta(\mathbf{p} - \mathbf{p}') [f (\mathbf{p},t) \delta_{s s'} - \mathbf{f}(\mathbf{p},t)\boldsymbol{\sigma}_{s s'}],
\label{eq:_B2} \\
    \langle 0, \text{in}|a^{\dagger}_{\mathbf{p},s}(t)b^{\dagger}_{-\mathbf{p'},s'}(t)|0, \text{in}\rangle &= \delta(\mathbf{p} - \mathbf{p}') [ \mathbf{u} (\mathbf{p},t) - i \mathbf{v} (\mathbf{p},t) ] \boldsymbol{\sigma}_{s' s},
\label{eq:_C2}\\
    \langle 0, \text{in}|b_{-\mathbf{p},s}(t)a_{\mathbf{p}',s'}(t)|0, \text{in}\rangle &=
    \delta(\mathbf{p} - \mathbf{p}') [ \mathbf{u} (\mathbf{p},t) + i \mathbf{v} (\mathbf{p},t) ] \boldsymbol{\sigma}_{s' s}.
\label{eq:_D2}
\end{align}
We already see that at asymptotically late times, the spin-summed electron density is given by $2f(\mathbf{p}, t_\text{out})$. The physical significance of the other QKE components is discussed in Sec.~\ref{sec:unrenorm}.
\end{widetext}

%%%%%%%%%%%%%%%%%%%%%%%%%%%%%%%%%%%%%%%%%%%%%%%%%%%%%%%%%

\section{Perturbation theory} \label{sec:PT}

Let us turn to the analysis of the QKEs. In this section, we will consider a weak-field regime and derive closed-form leading-order expressions for the QKE functions. To this end, we expand the coefficients of the system~\eqref{eq:qke_f}--\eqref{eq:qke_vvec} in powers of the field amplitude $E_0$, treating $\mathbf{E} (t)$ and $\mathbf{A} (t)$ as $\mathcal{O}(E_0)$ and formally taking $E_0 \to 0$. The definitions~\eqref{eq:omega_def}, \eqref{eq:mu1}, and \eqref{eq:mu2} yield
\begin{align}
\omega (\mathbf{p}, t) &= p_0 + \omega_1 (\mathbf{p}, t) + \ldots, \\
\boldsymbol{\mu}_1 (\mathbf{p}, t) &= \boldsymbol{\mu}_{1, 1} (\mathbf{p}, t) + \boldsymbol{\mu}_{1,2} (\mathbf{p}, t) + \ldots, \\
\boldsymbol{\mu}_2 (\mathbf{p}, t) &= \boldsymbol{\mu}_{2, 1} (\mathbf{p}, t) + \boldsymbol{\mu}_{2,2} (\mathbf{p}, t) + \ldots,
\end{align}
where the second index denotes the order in $E_0$. For example,
\begin{align}
\omega_1 (\mathbf{p}, t) &= -\frac{e \mathbf{A} (t) \mathbf{p}}{p_0}, \\
\boldsymbol{\mu}_{1,1} (\mathbf{p}, t) &= \frac{e}{2p_0 (p_0 + m)}\, [\mathbf{p} \times \mathbf{E} (t)],\\
\boldsymbol{\mu}_{2,1} (\mathbf{p}, t) &= \frac{e}{2p_0} \, \bigg \{ \frac{[\mathbf{p} \mathbf{E}(t)] \mathbf{p}}{p_0 (p_0 + m)} - \mathbf{E} (t) \bigg \}.
\end{align}
Power counting in Eqs.~\eqref{eq:qke_f}--\eqref{eq:qke_vvec} gives the leading orders of the QKE functions, motivating the following expansions:
\begin{align}
f (\mathbf{p}, t) &= f_2 (\mathbf{p}, t) + f_3 (\mathbf{p}, t) + \ldots \,,\\
\mathbf{f}(\mathbf{p}, t) &= \mathbf{f}_2 (\mathbf{p}, t) + \mathbf{f}_3 (\mathbf{p}, t) + \ldots \,,\\
\mathbf{u} (\mathbf{p}, t) &= \mathbf{u}_1 (\mathbf{p}, t) + \mathbf{u}_2 (\mathbf{p}, t) + \ldots \,,\\
\mathbf{v} (\mathbf{p}, t) &= \mathbf{v}_1 (\mathbf{p}, t) + \mathbf{v}_2 (\mathbf{p}, t) + \ldots
\end{align}
We now explicitly construct the leading-order contributions. To first order, Eqs.~\eqref{eq:qke_uvec} and \eqref{eq:qke_vvec} reduce to
\begin{align}
\dot{\mathbf{u}}_1 &= -\boldsymbol{\mu}_{2,1} + 2p_0\mathbf{v}_1, \\
\dot{\mathbf{v}}_1 &= -2p_0\mathbf{u}_1,
\end{align}
which yields
\begin{align}
\mathbf{u}_1 (\mathbf{p}, t) &= - \int\limits_{t_\text{in}}^{t} \! \boldsymbol{\mu}_{2,1} (\mathbf{p}, t') \cos[2p_0(t-t')] dt', \label{eq:tv_u_1}\\ 
\mathbf{v}_1 (\mathbf{p}, t) &= \int\limits_{t_\text{in}}^{t} \! \boldsymbol{\mu}_{2,1} (\mathbf{p}, t') \sin[2p_0(t-t')] dt'. \label{eq:tv_v_1}
\end{align}
Likewise, Eqs.~\eqref{eq:qke_f} and \eqref{eq:qke_fvec} give
\begin{align}
\dot{f}_2 &= -2 \boldsymbol{\mu}_{2,1} \mathbf{u}_1, \\
\dot{\mathbf{f}}_2 &= -2\boldsymbol{\mu}_{2,1} \times \mathbf{v}_1.
\end{align}
By integrating these equations, we obtain
\begin{align}
f_2 (\mathbf{p}, t) &= |\boldsymbol{\alpha} (\mathbf{p}, t) |^2, \\
\mathbf{f}_2 (\mathbf{p}, t) &= i [\boldsymbol{\alpha} (\mathbf{p}, t) \times \boldsymbol{\alpha}^* (\mathbf{p}, t)],
\end{align}
where
\begin{equation}
\boldsymbol{\alpha} (\mathbf{p}, t) = \int\limits_{t_\text{in}}^{t} \! \boldsymbol{\mu}_{2,1} (\mathbf{p}, t') \mathrm{e}^{2ip_0t'} dt'.
\end{equation}
Higher-order terms follow systematically from the recurrence relations
\begin{align}
\dot{f}_k &= -2 \sum_{l=1}^{k-1} \boldsymbol{\mu}_{2,k-l} \mathbf{u}_{l}, \\
\dot{\mathbf{f}}_k &= 2 \sum_{l=1}^{k-1} (\boldsymbol{\mu}_{1,k-l}\times\mathbf{f}_{l}) -2 \sum_{l=1}^{k-1} (\boldsymbol{\mu}_{2,k-l}\times\mathbf{v}_{l}), \\
\dot{\mathbf{u}}_k &= 2 \sum_{l=1}^{k-1} (\boldsymbol{\mu}_{1,k-l}\times\mathbf{u}_{l}) + 2 \sum_{l=1}^{k-1} \boldsymbol{\mu}_{2,k-l} f_{l} \nonumber \\
{}&- \boldsymbol{\mu}_{2,k} + 2 p_0 \mathbf{v}_{k} + 2 \sum_{l=1}^{k-1} \omega_{k-l} \mathbf{v}_{l}, \\
\dot{\mathbf{v}}_k &= 2 \sum_{l=1}^{k-1} (\boldsymbol{\mu}_{1,k-l}\times\mathbf{v}_{l}) - 2 \sum_{l=1}^{k-1} (\boldsymbol{\mu}_{2,k-l}\times\mathbf{f}_{l}) \nonumber\\
{}&- 2 p_0 \mathbf{u}_{k} - 2 \sum_{l=1}^{k-1} \omega_{k-l} \mathbf{u}_{l}
\end{align}
for $k = 2$, $3$, ... We assume here that $f_1 = \mathbf{f}_1 = 0$. Note that in the case of a linearly polarized external field, a perturbative hierarchy was recently examined within a different kinetic-theory framework~\cite{edwards_prdl_2025}.

In what follows, the leading-order terms will be used within the renormalization procedure.

%%%%%%%%%%%%%%%%%%%%%%%%%%%%%%%%%%%%%%%%%%%%%%%%%%%%%%%%%

\section{Unrenormalized current, energy-momentum tensor, and angular-momentum tensor} \label{sec:unrenorm}

In this section, we will compute the main dynamical quantities in terms of the QKE functions $f$, $\mathbf{f}$, $\mathbf{u}$, and $\mathbf{v}$. These expressions will also clarify the physical meaning of the kinetic functions.

\subsection{Electron-positron current}

We continue working in the Heisenberg representation, and we evaluate the time-dependent mean value
\begin{equation}
j^\mu (t) = \frac{e}{2} \frac{1}{V} \int \! d\mathbf{x} \, \big \langle 0, \text{in} \big | \big [\bar{\psi}(x), \, \gamma^\mu \psi(x) \big ] \big | 0, \text{in} \big \rangle.
\label{eq:j_mean}
\end{equation}
First, we plug in the expansion~\eqref{eq:psi_adiabatic} and carry out the spatial integration taking into account the explicit form of the adiabatic functions~\eqref{eq:adiabatic_plus} and \eqref{eq:adiabatic_minus}. Second, the in-vacuum expectation values then yield the correlation functions that have already been found in Eqs.~\eqref{eq:_A2}--\eqref{eq:_D2}. Finally, the bispinor parts of the form $\overline{u}_{\mathbf{p} - e\mathbf{A}(t),s} \gamma^\mu u_{\mathbf{p} - e\mathbf{A}(t),s'}$ are evaluated by means of the following relations:
\begin{align}
\overline{u}_{\mathbf{p},s} \boldsymbol{\gamma} u_{\mathbf{p},s'} &= \frac{\mathbf{p}}{p^0} \, \delta_{s s'}, \\
\overline{v}_{\mathbf{p},s} \boldsymbol{\gamma} v_{\mathbf{p},s'} &= -\frac{\mathbf{p}}{p^0} \, \delta_{s s'}, \\
\overline{u}_{\mathbf{p},s} \boldsymbol{\gamma} v_{\mathbf{p},s'} &= \overline{v}_{\mathbf{p},s} \boldsymbol{\gamma} u_{\mathbf{p},s'} \nonumber \\
{}&= \boldsymbol{\sigma}_{s s'} - \frac{\mathbf{p} (\boldsymbol{\sigma}_{s s'} \mathbf{p})}{p^0 (p^0 + m)}.
\end{align}
Since the diagonal parts ($s=s'$) of Eqs.~\eqref{eq:_A2} and \eqref{eq:_B2} coincide, the scalar part of the current vanishes, $j^0 (t) = 0$. This result indicates that the effects of pair production and vacuum polarization preserve the electric neutrality of the system. The spatial part of $j^\mu (t)$ can be written as
\begin{equation}
\mathbf{j} (t) = \mathbf{j}_\text{cond} (t) + \mathbf{j}_\text{pol} (t),
\label{eq:j_total_unrenorm}
\end{equation}
with conduction and polarization parts
\begin{align}
\mathbf{j}_\text{cond} (t) &= 4e \int \! \frac{d\mathbf{p}}{(2\pi)^3} \, \frac{\mathbf{q}}{\omega} \, f (\mathbf{p},t), \label{eq:jcond} \\
\mathbf{j}_\text{pol} (t) &= 4e \int \! \frac{d\mathbf{p}}{(2\pi)^3} \bigg \{ \mathbf{u} (\mathbf{p},t) - \frac{[\mathbf{q} \mathbf{u}(\mathbf{p},t)]\mathbf{q}}{\omega (\omega + m)} \bigg \}. \label{eq:jpol}
\end{align}
Here we recall that $\mathbf{q} = \mathbf{p} - e \mathbf{A} (t)$ depends on $t$ and $\omega = \sqrt{m^2 + \mathbf{q}^2} \equiv q^0$. The above expressions coincide with the results of Refs.~\cite{BB_prd_1991,hebenstreit_prd_2010}, where the current density was evaluated within the WKB formalism. To make contact with the DHW formulation, one may use the relations deduced in Ref.~\cite{aleksandrov_kudlis_klochai} and presented in Appendix~\ref{app:BB}. Note that at intermediate times, a clean separation between real and virtual contributions is not available, so the conduction and polarization currents admit this interpretation only for $t \geqslant t_\text{out}$. However, their sum $\mathbf{j} (t)$ is well defined by virtue of N\"other's theorem. We also underline that within the Heisenberg representation, the in-vacuum state $|0, \text{in} \rangle$ is time-independent, so Eq.~\eqref{eq:j_mean} therefore gives the full time dependence of the total current within the external-field approximation (i.e., neglecting radiative corrections). We also emphasize that Eqs.~\eqref{eq:jcond} and \eqref{eq:jpol} are gauge-invariant. Any constant shift of the vector potential $\mathbf{A} (t)$ can be absorbed by shifting the integration variable~$\mathbf{p}$ [note that the coefficients of the QKEs~\eqref{eq:qke_f}--\eqref{eq:qke_vvec} depend only on the kinetic momentum~$\mathbf{q}$].

As will be seen below, the integral in Eq.~\eqref{eq:jcond} converges, whereas the $\mathbf{p}$ integration in Eq.~\eqref{eq:jpol} is logarithmically divergent. To demonstrate this, we will analyze the large-momentum behavior of the solutions of the QKE system~\eqref{eq:qke_f}--\eqref{eq:qke_vvec}. However, let us first evaluate the unrenormalized energy-momentum tensor.

\subsection{Energy-momentum tensor}

To calculate the expectation values of the energy-momentum tensor $T^{\mu \nu}$, we will separately consider its different elements. First, the $T^{00}$ component represents the energy per unit volume~\cite{greiner_sfqed}:
\begin{equation}
T^{00} (t) = \frac{1}{2} \frac{1}{V} \int \! d\mathbf{x} \, \big \langle 0, \text{in} \big | \big [\psi^\dagger (x), \mathcal{H}_e (t) \psi (x) \big ] \big | 0, \text{in} \big \rangle.
\end{equation}
The $T^{0k}$ elements yield the momentum density:
\begin{equation}
T^{0k} (t) = \frac{i}{2} \frac{1}{V} \int \! d\mathbf{x} \, \big \langle 0, \text{in} \big | \big [\psi^\dagger (x), D^k \psi (x) \big ] \big | 0, \text{in} \big \rangle.
\label{eq:T0k_start}
\end{equation}
where $D^k = \partial^k + ieA^k (t)$. The remaining elements, i.e., stress components, are given by
\begin{equation}
T^{kl} (t) = \frac{i}{4} \frac{1}{V} \int \! d\mathbf{x} \, \big \langle 0, \text{in} \big | \big [\psi^\dagger (x), \mathcal{T}^{kl} \psi (x) \big ] \big | 0, \text{in} \big \rangle,
\end{equation}
where $\mathcal{T}^{kl}=\gamma^0( \gamma^k D^l + \gamma^l D^k)$. In our calculations, we drop the field-independent vacuum contributions that appear in the diagonal components of $T^{\mu \nu}$.

First, we find
\begin{equation}
T^{00} (t) = 4 \int \! \frac{d\mathbf{p}}{(2\pi)^3} \, \omega (\mathbf{p},t) f (\mathbf{p},t), \label{eq:T00}
\end{equation}
As in Eqs.~\eqref{eq:jcond} and \eqref{eq:jpol}, the factor of $4$ appears due to summation over particles ($e^+$ and $e^-$) and over spin states. One can also formally show that
\begin{equation}
\dot{T}^{00} (t) = \mathbf{E} (t) \mathbf{j} (t)
\label{eq:energy_current}
\end{equation}
in accordance with the expected work-energy relation in the external-field approximation. As will be discussed below, the momentum integrals in the energy density~\eqref{eq:T00} and the polarization current~\eqref{eq:jpol} logarithmically diverge. We will perform renormalization of these quantities and show that the relation~\eqref{eq:energy_current} continues to hold also for the renormalized (finite) functions.

Second, the total momentum is zero,
\begin{equation}
T^{0k} (t) = 0. \label{eq:T0k}
\end{equation}
This result holds also for the total generalized momentum, which contains $\partial^k$ instead of $D^k$ in Eq.~\eqref{eq:T0k_start}. The external field does not change the generalized momentum of the initial vacuum state, nor does it transfer any kinetic momentum as the electrons and positrons are accelerated in opposite directions.

Finally, we obtain
\begin{align}
T^{kl} (t) &= 4 \int \! \frac{d\mathbf{p}}{(2\pi)^3} \, \bigg \{ \frac{q^k q^l}{\omega} \, \bigg [ f (\mathbf{p},t) - \frac{\mathbf{q} \mathbf{u} (\mathbf{p},t)}{\omega + m}\bigg ] \nonumber \\
{}&+ \frac{1}{2} \, \big [ q^k u^l (\mathbf{p},t) + q^l u^k (\mathbf{p},t) \big ] \bigg \}. \label{eq:Tkl}
\end{align}
The sum of the diagonal elements reads
\begin{align}
T^{kk} (t) &= 4 \int \! \frac{d\mathbf{p}}{(2\pi)^3} \, \bigg \{ \frac{\mathbf{q}^2}{\omega} \, f (\mathbf{p},t) + \frac{m}{\omega} \, [\mathbf{q} \mathbf{u} (\mathbf{p},t) ] \bigg \}. \label{eq:Tkk}
\end{align}
As for the current density, the term involving $f (\mathbf{p},t)$ and coming from the normal correlations functions~\eqref{eq:_A2} and \eqref{eq:_B2} can be associated with real particles, whereas the terms with $\mathbf{u} (\mathbf{p},t)$ originating from the anomalous correlators~\eqref{eq:_C2} and \eqref{eq:_D2} describe the effects of virtual pairs. On the other hand, at intermediate times, these two contributions cannot be unambiguously disentangled.

\subsection{Angular-momentum tensor}

We will next calculate the angular-momentum tensor:
\begin{equation}
M^{\mu\nu} (t) = \frac{i}{2} \frac{1}{V} \int \! d\mathbf{x} \, \big \langle 0, \text{in} \big | \big [\psi^\dagger (x), \mathcal{M}^{\mu\nu} \psi (x) \big ] \big | 0, \text{in} \big \rangle,
\end{equation}
where $\mathcal{M}^{\mu\nu} = x^\mu D^\nu - x^\nu D^\mu - (i/2) \sigma^{\mu \nu}$ and $\sigma^{\mu \nu} = (i/2) [\gamma^\mu, \, \gamma^\nu]$. First, we note that the orbital part $x^\mu D^\nu - x^\nu D^\mu$ does not yield any nonzero contributions due to the spatial homogeneity of the external field. Second, the spin part can be computed with the aid of the following relations:
\begin{align}
u^\dagger_{\mathbf{p},s} \boldsymbol{\Sigma} u_{\mathbf{p},s'} &= v^\dagger_{\mathbf{p},s} \boldsymbol{\Sigma} v_{\mathbf{p},s'} \nonumber \\
{}&= \frac{m}{p^0} \, \boldsymbol{\sigma}_{s s'} +  \frac{\mathbf{p} (\boldsymbol{\sigma}_{s s'} \mathbf{p})}{p^0 (p^0 + m)}, \\
u^\dagger_{\mathbf{p},s} \boldsymbol{\Sigma} v_{\mathbf{p},s'} &= - v^\dagger_{\mathbf{p},s} \boldsymbol{\Sigma} u_{\mathbf{p},s'} = -\frac{i}{p^0} \, (\mathbf{p} \times \boldsymbol{\sigma}_{s s'}),
\end{align}
where $\boldsymbol{\Sigma}$ is conventionally defined via $\sigma^{kl} = \varepsilon^{kln} \Sigma^n$. In terms of the spin vector $\mathbf{S}$ ($M^{kl} = \varepsilon^{kln} S^n$), we obtain:
\begin{align}
\mathbf{S} (t) &= 2 \int \! \frac{d\mathbf{p}}{(2\pi)^3} \, \bigg \{ \frac{m}{\omega} \, \mathbf{f} (\mathbf{p},t) - \frac{\mathbf{q} \times \mathbf{v} (\mathbf{p},t)}{\omega} \nonumber \\
{}&+ \frac{[\mathbf{q} \mathbf{f}(\mathbf{p},t)]\mathbf{q}}{\omega (\omega + m)} \bigg \}. \label{eq:Spin}
\end{align}
This result is in full agreement with Ref.~\cite{BB_prd_1991}.

According to Eq.~\eqref{eq:_A2}, the final spin-resolved number density~\eqref{eq:np_A} of the electrons produced is given by
\begin{equation}
\frac{(2\pi)^3}{V} \, n_{\mathbf{p},s}^{(e^-)} = f (\mathbf{p}, t_\text{out}) + (\mathrm{sign} \, s) f_z (\mathbf{p}, t_\text{out}).
\label{eq:np_el}
\end{equation}
Since our gauge choice corresponds to $\mathbf{A}_\text{out} = 0$, at asymptotically late times, the generalized momentum and the kinetic one coincide, $\mathbf{p} = \mathbf{q}(t_\text{out})$. As was discussed in detail in Ref.~\cite{aleksandrov_kudlis_klochai}, Eq.~\eqref{eq:np_el} yields the mean number of the electrons in a given spin state $s$ described by the specific bispinors~\eqref{eq:u_explicit_1} and \eqref{eq:u_explicit_2} chosen as a basis within the above derivations. In other words, the form of $n_{\mathbf{p},s}^{(e^-)}$ is sensitive to the choice of the bispinor basis. For instance, one may instead classify particles by helicity, i.e., the spin projection onto the propagation direction. In this case, the basis bispinors~\eqref{eq:u_explicit_1} and \eqref{eq:u_explicit_2} should be unitary transformed, so that the new bispinors are eigenvectors of the operator $(\boldsymbol{\Sigma} \mathbf{p})/|\mathbf{p}|$ (with eigenvalues $\pm 1$). The helicity-resolved densities of the electrons take the following form~\cite{aleksandrov_kudlis_klochai}:
\begin{align}
\frac{(2\pi)^3}{V} n^{(e^-\text{L})}_\mathbf{p} &= f(\mathbf{p}, t_\text{out}) - \frac{\mathbf{q} \mathbf{f} (\mathbf{p}, t_\text{out})}{|\mathbf{q}|}, \label{eq:n_eL}\\
\frac{(2\pi)^3}{V} n^{(e^-\text{R})}_\mathbf{p} &= f(\mathbf{p}, t_\text{out}) + \frac{\mathbf{q} \mathbf{f} (\mathbf{p}, t_\text{out})}{|\mathbf{q}|}, \label{eq:n_eR}
\end{align}
where ``L'' and ``R'' stand for the left-handed (negative) and right-handed (positive) helicity, respectively. Note that the spin-summed densities are determined solely by $f(\mathbf{p}, t_\text{out})$ and are basis-independent. For the positrons produced, the analogous densities read
\begin{align}
n^{(e^+\text{L})}_\mathbf{p} &= n^{(e^-\text{R})}_{-\mathbf{p}}, \label{eq:n_pL} \\
n^{(e^+\text{R})}_\mathbf{p} &= n^{(e^-\text{L})}_{-\mathbf{p}}, \label{eq:n_pR}
\end{align}
From Eqs.~\eqref{eq:np_el}--\eqref{eq:n_eR}, it is clear that the function $\mathbf{f}$ describes the spin effects in the final particle distributions. The term with $\mathbf{v}$ involved in Eq.~\eqref{eq:Spin} originates from the anomalous correlation functions~\eqref{eq:_C2} and \eqref{eq:_D2} and can be interpreted as a vacuum-polarization contribution to the spin density induced by virtual pairs. The final densities~\eqref{eq:n_eL}--\eqref{eq:n_pR} of the real particles to be detected do not contain $\mathbf{v}$.

\subsection{Summary. Physical interpretation of the QKE functions}

The functions $f$, $\mathbf{f}$, $\mathbf{u}$, and $\mathbf{v}$ are coupled by the QKEs~\eqref{eq:qke_f}--\eqref{eq:qke_vvec}. They evolve in time, so that the following integral of motion is always preserved~\cite{aleksandrov_kudlis_klochai}:
\begin{equation}
\frac{1}{4} (1-2f)^2 + \mathbf{f}^2 + \mathbf{u}^2 + \mathbf{v}^2 = \frac{1}{4}.
\label{eq:qke_int_of_motion_gen}
\end{equation}
This conservation law is related to the {\it unitary} evolution of the one-particle solutions of the Dirac equation.

The QKE functions describe different physical effects, which can be interpreted based on the results of the previous subsection. The overall dynamics of the real and virtual particles corresponding to the conduction and polarization effects, respectively, is described by the functions $f$ and $\mathbf{u}$. The dynamics of the spin degrees of freedom in these two parts is governed by the functions $\mathbf{f}$ and $\mathbf{v}$, respectively.

Within the DHW formalism, the physical interpretation of the kinetic functions was discussed in Refs.~\cite{BB_prd_1991,hebenstreit_prd_2010}. The DHW functions are explicitly related to the QKE components as was demonstrated in Ref.~\cite{aleksandrov_kudlis_klochai} (see Appendix~\ref{app:BB}). To gain a deeper understanding of the physics, it is instructive to transition between the DHW and QKE descriptions. For example, as was already mentioned in Sec.~\ref{sec:derivation}, the function $f$ represents the spin-summed phase-space density of the particles. According to Eqs.~\eqref{eq:jcond} and \eqref{eq:jpol}, the analogous density of the current is given by a nontrivial combination of $f$ and $\mathbf{u}$. In the DHW approach, this combination directly corresponds to the function $\tilde{\boldsymbol{\mathfrak{v}}}$ [see Eq.~\eqref{eq:rel_v}], which was accordingly interpreted in Refs.~\cite{BB_prd_1991,hebenstreit_prd_2010}.

\subsection{Large-momentum behavior and ultraviolet divergences} \label{sec:large_momentum}

To examine the convergence of the $\mathbf{p}$ integrals derived above, we inspect the large-$\mathbf{p}$ asymptotic behavior of the QKE functions. To this end, we first expand the coefficients in the system~\eqref{eq:qke_f}--\eqref{eq:qke_vvec}:
\begin{align}
\omega(\mathbf{p}, t) &= |\mathbf{p}| - e \mathbf{A}(t) \mathbf{n}+ \mathcal{O} \bigg ( \frac{1}{|\mathbf{p}|} \bigg ), \\
\boldsymbol{\mu}_1 (\mathbf{p}, t) &= \frac{\mathbf{n} \times e \mathbf{E}(t)}{2|\mathbf{p}|} + \mathcal{O} \bigg ( \frac{1}{|\mathbf{p}|^2} \bigg ), \\
\boldsymbol{\mu}_2 (\mathbf{p}, t) &= \frac{\mathbf{n} [e \mathbf{E}(t) \mathbf{n}]}{2|\mathbf{p}|} - \frac{e\mathbf{E}(t)}{2|\mathbf{p}|} + \mathcal{O} \bigg ( \frac{1}{|\mathbf{p}|^2} \bigg ),
\end{align}
where $\mathbf{n} = \mathbf{p}/|\mathbf{p}|$. The solutions can be constructed to the required order as in Sec.~\ref{sec:PT}:
\begin{align}
f (\mathbf{p}, t) &= \frac{e^2}{16|\mathbf{p}|^4} \, \Big \{ \mathbf{E}^2 (t) - [\mathbf{n} \mathbf{E}(t)]^2 \Big \}  + \mathcal{O} \bigg ( \frac{1}{|\mathbf{p}|^5} \bigg ), \label{eq:large_p_f}\\
\mathbf{f} (\mathbf{p}, t) &= \frac{e^2}{16 |\mathbf{p}|^5} \, \Big \{ \mathbf{E}(t) \times \dot{\mathbf{E}}(t) + [\mathbf{n} \dot{\mathbf{E}}(t)] [\mathbf{n} \times \mathbf{E}(t)] \nonumber \\
{}&- [\mathbf{n} \mathbf{E}(t)] [\mathbf{n} \times \dot{\mathbf{E}}(t)] \Big \} + \mathcal{O} \bigg ( \frac{1}{|\mathbf{p}|^6} \bigg ), \label{eq:fvec_large_momentum} \\
\mathbf{u} (\mathbf{p}, t) &= \frac{e\dot{\mathbf{E}}(t) - \mathbf{n} [e \dot{\mathbf{E}}(t) \mathbf{n}]}{8 |\mathbf{p}|^3} + \mathcal{O} \bigg ( \frac{1}{|\mathbf{p}|^4} \bigg ), \\
\mathbf{v} (\mathbf{p}, t) &= -\frac{e\mathbf{E}(t) - \mathbf{n} [e \mathbf{E}(t) \mathbf{n}]}{4 |\mathbf{p}|^2} + \mathcal{O} \bigg ( \frac{1}{|\mathbf{p}|^3} \bigg ). \label{eq:large_p_vvec}
\end{align}
Note that the expression in Eq.~\eqref{eq:fvec_large_momentum} vanishes unless the external field rotates in space. This is in accordance with the fact that the spin effects are absent in the case of a linearly polarized electric background. It is now evident that the polarization current~\eqref{eq:jpol}, energy~\eqref{eq:T00}, and stress tensor~\eqref{eq:Tkl} contain logarithmically divergent integrals [any would-be linear divergences in Eq.~\eqref{eq:Tkl} cancel after angular integration]. Therefore, these quantities require renormalization. 

Note, however, that for $t>t_\text{out}$ the $\mathbf{p}$ integrals converge and provide finite results [the leading-order power expansions in Eqs.~\eqref{eq:large_p_f}--\eqref{eq:large_p_vvec} vanish at asymptotically late times]. In the case of a linearly polarized external field, this conclusion was drawn in Ref.~\cite{GMM}. Moreover, it turns out that the finite results for $t>t_\text{out}$ do not require any additional (finite) subtractions, i.e., at asymptotic times they already provide physically relevant predictions. To see this and to obtain finite physical results at intermediate times $t_\text{in} < t < t_\text{out}$, we will perform careful charge renormalization. In the next section, we will discuss in detail two approaches to constructing the renormalized current density and also derive finite expressions for the renormalized energy-momentum tensor.

%%%%%%%%%%%%%%%%%%%%%%%%%%%%%%%%%%%%%%%%%%%%%%%%%%%%%%%%%

\section{Renormalization} \label{sec:renorm}

Here we will describe how the logarithmic divergences identified above are removed by charge renormalization. We will consider the polarization current~\eqref{eq:jpol} and employ two complementary methods: (i) renormalizing the one-potential contribution and (ii) procedure based on the Pauli-Villars regularization. After that, we will obtain the renormalized energy-momentum tensor.

First, one can extract the leading-order contributions with respect to the external field strength $\mathbf{E} (t)$ and subtract the divergent parts ensuring that the subtraction has the form of standard counterterms in the QED Lagrangian (or effective action). In this case, one has to carefully verify that the subtraction procedure is implemented in a gauge-invariant manner: the finite results upon subtraction may still contain spurious {\it finite} terms. Another approach is more systematic and based on manifestly gauge-invariant Pauli-Villars regularization~\cite{pauli_villars_1949}. In this case, one has to analyze the large-electron-mass behavior of the unrenormalized quantities and subtract the divergent parts (and any associated spurious finite terms) and then take $M \to \infty$. The spurious terms, if any, will be subtracted automatically, and one can explicitly see that the subtraction contributions have the form of the initial Lagrangian terms. In what follows, we will employ both of the above techniques.

\subsection{One-potential contribution to the current density and subtraction term}

We first extract the leading-order contribution to the polarization current~\eqref{eq:jpol} with respect to the interaction with the external field [the finite conduction term~\eqref{eq:jcond} has no linear term in the field amplitude and starts at the third order]. By means of Eq.~\eqref{eq:tv_u_1}, we immediately obtain
\begin{align}
\mathbf{j}^{(\text{1P})} (t) &= 2e^2 \int \! \frac{d\mathbf{p}}{(2\pi)^3} \frac{1}{p_0} \int\limits_{t_\text{in}}^{t} \! dt' \, \bigg \{ \mathbf{E} (t') - \frac{[\mathbf{p} \mathbf{E}(t')]\mathbf{p}}{p_0^2} \bigg \} \nonumber \\
{}&\times \cos[2p_0(t-t')]. \label{eq:jpol_1P}
\end{align}
Since this term is linear in $\mathbf{E}$, we refer to it as the one-potential contribution. One can average here over the angles by means of the identity $\langle p^i p^j \rangle = \mathbf{p}^2 \delta^{ij}/3$:
\begin{align}
\mathbf{j}^{(\text{1P})} (t) &= \frac{4e^2}{3} \int \! \frac{d\mathbf{p}}{(2\pi)^3} \frac{1}{p_0} \bigg ( 1 + \frac{m^2}{2p_0^2} \bigg ) \nonumber \\
{}&\times \int\limits_{t_\text{in}}^{t} \! dt' \, \mathbf{E} (t') \cos[2p_0(t-t')]. \label{eq:jpol_1P_angles}
\end{align}
Let us now formally integrate by parts in $t'$ (we always assume that the external field profile is sufficiently smooth):
\begin{align}
\mathbf{j}^{(\text{1P})} (t) &= \frac{e^2}{3} \, \dot{\mathbf{E}} (t) \int \! \frac{d\mathbf{p}}{(2\pi)^3} \frac{1}{p_0^3} \bigg ( 1 + \frac{m^2}{2p_0^2} \bigg ) \nonumber \\
{}& - \frac{e^2}{3} \int \! \frac{d\mathbf{p}}{(2\pi)^3} \frac{1}{p_0^3} \bigg ( 1 + \frac{m^2}{2p_0^2} \bigg ) \nonumber \\
{}& \times \int\limits_{t_\text{in}}^{t} \! dt' \, \ddot{\mathbf{E}} (t') \cos[2p_0(t-t')]. \label{eq:jpol_1P_2}
\end{align}
Whereas the second term is finite, the first one diverges logarithmically. We will now demonstrate that it is the second (finite) contribution that corresponds to the {\it renormalized} one-potential current density $\mathbf{j}^{(\text{1P})}_\text{R} (t)$, while the first term in Eq.~\eqref{eq:jpol_1P_2} should be discarded (it is absorbed into the charge renormalization constant). To this end, we will first note that the external electric field is generated by the external current $\mathbf{j}_\text{ext} (t) = - \dot{\mathbf{E}} (t)$, $j^0_\text{ext} (t) = 0$:
\begin{equation}
A_\mu (x) = \int \limits_{t_\text{in}}^{x^0} dy^0 \int d\mathbf{y} D^{(\text{ret})}_{\mu \nu} (x-y) j^\nu_\text{ext} (y),
\label{eq:A_j_ext}
\end{equation}
where $D^{(\text{ret})}_{\mu \nu} (x)$ is the retarded photon Green's function:
\begin{equation}
D^{(\text{ret})}_{\mu \nu} (x) = g_{\mu \nu} \theta (x^0) \int \frac{d\mathbf{k}}{(2\pi)^3} \, \mathrm{e}^{i\mathbf{k} \mathbf{x}} \, \frac{\sin |\mathbf{k}|x^0}{|\mathbf{k}|}.
\end{equation}
The representation~\eqref{eq:A_j_ext} corresponds to the gauge $\mathbf{A}_\text{in} = 0$ (our results will be gauge independent). One can verify it by a direct evaluation of the right-hand side:
\begin{equation}
\mathbf{A} (t) = - \int \limits_{t_\text{in}}^{t} \mathbf{E} (t') dt',\quad A_0 = 0.
\end{equation}
To first order in $\alpha$ and first order in the external field, the correction to $\mathbf{A} (t)$ can be obtained by means of the renormalized polarization tensor via
\begin{align}
A^{(\text{1P})}_\mu (x) &= \int \limits_{t_\text{in}}^{x^0} dy^0 \int d\mathbf{y} \int\limits_{\mathcal{C}} \frac{d^4k}{(2\pi)^4} \, \mathrm{e}^{-ik(x-y)} \, j^\nu_\text{ext} (y) \nonumber \\
{}& \times \frac{k_\mu k_\nu -k^2g_{\mu \nu}}{(k^2 - m_\gamma^2)^2} \, \Pi_\text{R} (k^2),
\label{eq:A1_j_ext}
\end{align}
where the contour $\mathcal{C}$ in the $k^0$ complex plane encompasses both poles and has a clockwise direction. The auxiliary photon mass $m_\gamma$ should be replaced by zero at the end of the calculations. It is convenient to exploit here the dispersion relation for the renormalized polarization function~\cite{blp}
\begin{equation}
\Pi_\text{R} (k^2) = \frac{k^2}{\pi} \int \limits_{4m^2}^{\infty} dz \, \frac{\mathrm{Im} \, \Pi_\text{R} (z)}{z(z-k^2 - i \varepsilon)},
\label{eq:Pi_disp}
\end{equation}
where
\begin{align}
\mathrm{Im} \, \Pi_\text{R} (z) &= - \frac{e^2}{12 \pi} \bigg ( 1 + \frac{2m^2}{z} \bigg ) \nonumber \\
{}& \times \sqrt{1 - 4m^2/z} \, \theta (z - 4m^2).
\end{align}
In our setup, by integrating over $\mathbf{y}$, one obtains $\delta (\mathbf{k})$, so the nontrivial integrations are those over $y^0 \equiv t'$, $k^0$, and $z$. Due to the prefactor $k^2 = k_0^2$ in Eq.~\eqref{eq:Pi_disp}, one can immediately set $m_\gamma = 0$. The $k^0$ integration can be easily performed by means of the residue theorem: there are two contributions from $k_0 = \pm \sqrt{z}$. With the aid of the relation
\begin{equation}
\int \limits_{m}^{\infty} f(w) dw = \frac{1}{4\pi} \int d\mathbf{p} \, \frac{f(p_0)}{p_0 \sqrt{p_0^2 - m^2}}, 
\end{equation}
we arrive at
\begin{align}
\mathbf{A}^{(\text{1P})} (t) &= - \frac{e^2}{6} \int \! \frac{d\mathbf{p}}{(2\pi)^3} \frac{1}{p_0^4} \bigg ( 1 + \frac{m^2}{2p_0^2} \bigg ) \nonumber \\
{}& \times \int\limits_{t_\text{in}}^{t} \! dt' \, \dot{\mathbf{E}} (t') \sin[2p_0(t-t')]. \label{eq:A1}
\end{align}
In terms of the field strength $\mathbf{E}^{(1)} (t) = - \dot{\mathbf{A}}^{(1)} (t)$, we find
\begin{align}
\mathbf{E}^{(\text{1P})} (t) &= \frac{e^2}{3} \int \! \frac{d\mathbf{p}}{(2\pi)^3} \frac{1}{p_0^3} \bigg ( 1 + \frac{m^2}{2p_0^2} \bigg ) \nonumber \\
{}& \times \int\limits_{t_\text{in}}^{t} \! dt' \, \dot{\mathbf{E}} (t') \cos[2p_0(t-t')]. \label{eq:E1}
\end{align}
This induced field is the time-dependent analogue of the Uehling correction~\cite{blp,uehling}. The above expression is finite and obtained by the conventional charge renormalization procedure. The current density in the leading order with respect to $\mathbf{E} (t)$ is now given by $\mathbf{j}^{(\text{1P})}_\text{R} (t) = -\dot{\mathbf{E}}^{(\text{1P})} (t)$ and reads
\begin{align}
\mathbf{j}^{(\text{1P})}_\text{R} (t) &= -\frac{e^2}{3} \int \! \frac{d\mathbf{p}}{(2\pi)^3} \frac{1}{p_0^3} \bigg ( 1 + \frac{m^2}{2p_0^2} \bigg ) \nonumber \\
{}& \times \int\limits_{t_\text{in}}^{t} \! dt' \, \ddot{\mathbf{E}} (t') \cos[2p_0(t-t')]. \label{eq:j1R}
\end{align}
This expression indeed coincides with the second (finite) term of Eq.~\eqref{eq:jpol_1P_2}.

\begin{figure*}[t]
    \centering
    \includegraphics[width=0.8\linewidth]{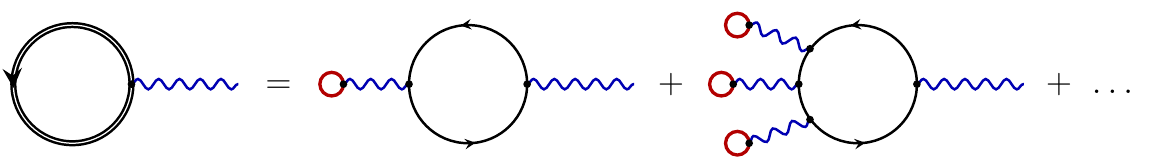}
    \caption{Series representation of the one-loop current density~\eqref{eq:j_mean}. The interaction with the external background is viewed as a one-photon exchange between the vacuum fermionic loop (black) and the classical external current $\mathbf{j}_\text{ext} (t) = -\dot{\mathbf{E}} (t)$ (red small circles). The first diagram in the right-hand side encodes the one-potential contribution $\mathbf{j}^{(\text{1P})} (t) = \mathbf{j}^{(\text{1P})}_\text{pol} (t)$. The diagrams with an even number of the external-field legs vanish according to Furry's theorem.}
    \label{fig:diag_series}
\end{figure*}

The full renormalized polarization current then follows from subtracting the unrenormalized one-potential contribution~\eqref{eq:jpol_1P} and adding the renormalized one-potential term~\eqref{eq:j1R}:
\begin{equation}
\mathbf{j}_\text{pol,R} (t) = \mathbf{j}_\text{pol} (t) - \mathbf{j}^{(\text{1P})} (t) + \mathbf{j}^{(\text{1P})}_\text{R} (t),
\end{equation}
which can be written explicitly as
\begin{align}
\mathbf{j}_\text{pol,R} (t) &= 4e \int \! \frac{d\mathbf{p}}{(2\pi)^3} \bigg \{ \mathbf{u} (\mathbf{p},t) - \frac{[\mathbf{q} \mathbf{u}(\mathbf{p},t)]\mathbf{q}}{\omega (\omega + m)} \nonumber \\
{}& - \frac{1}{12} \frac{1}{p_0^3} \bigg ( 1 + \frac{m^2}{2p_0^2} \bigg ) e\dot{\mathbf{E}} (t) \bigg \} .\label{eq:jpol_R}
\end{align}
Equivalently, one may write
\begin{equation}
\mathbf{j}_\text{pol,R} (t) = 4e \int \! \frac{d\mathbf{p}}{(2\pi)^3} \bigg \{ \mathbf{u}_\text{R} (\mathbf{p},t) - \frac{[\mathbf{q} \mathbf{u}_\text{R} (\mathbf{p},t)]\mathbf{q}}{\omega (\omega + m)} \bigg \},
\label{eq:jpol_R_uR}
\end{equation}
where $\mathbf{u}_\text{R} = \mathbf{u} - \delta \mathbf{u}$ and
\begin{equation}
\delta \mathbf{u} (\mathbf{p},t) \equiv \frac{e}{8 \omega^3} \, \bigg \{ \dot{\mathbf{E}}(t) - \frac{[\mathbf{q} \dot{\mathbf{E}}(t)] \mathbf{q}}{\omega (\omega + m)} \bigg \}.
\label{eq:uR}
\end{equation}
The conduction current~\eqref{eq:jcond} does not require any renormalization. The above calculations can be illustrated by the diagrams depicted in Fig.~\ref{fig:diag_series}. The first-order diagram yields the one-potential contribution to the current density $\mathbf{j}^{(\text{1P})} (t) = \mathbf{j}^{(\text{1P})}_\text{pol} (t)$, which after renormalization of the polarization tensor takes the form~\eqref{eq:j1R}. Note that the second-order contribution to the conduction current~\eqref{eq:jcond} vanishes since the integrand is odd in $\mathbf{p}$, consistent with Furry's theorem.

It is well known that in the external Coulomb field induced by a motionless charge, e.g. a nucleus, the analogous calculations require additional care because the final finite expression still contains a spurious contribution of the third order in the external field~\cite{gyulassy_1975,rinker_wilets_1975,borie_rinker_1982,soff_mohr_1988} (it is a part of the unrenormalized second term in the right-hand side in Fig.~\ref{fig:diag_series}). This spurious contribution can be extracted by replacing the electron mass $m$ with $\lambda m$ and taking the limit $\lambda \to \infty$ and must be subtracted as well. As was shown in Ref.~\cite{aleksandrov_prd_2019_2}, no such spurious terms arise in the present setup, so the prescriptions~\eqref{eq:jpol_R_uR} and \eqref{eq:uR} yield the correct result.

To finalize the present discussion, let us now provide a simplified expression for the Uehling-like contribution~\eqref{eq:E1}:
\begin{align}
\mathbf{E}^{(\text{1P})} (t) &= \frac{2\alpha}{3\pi} \int\limits_{1}^{\infty} dz \, \frac{\sqrt{z^2 - 1}}{z^2} \bigg ( 1 + \frac{1}{2z^2} \bigg ) \nonumber \\
{}& \times \int\limits_{t_\text{in}}^{t} \! dt' \, \dot{\mathbf{E}} (t') \cos[2mz(t-t')]. \label{eq:E1_simple}
\end{align}

In what follows, we will show how to obtain the finite renormalized current~\eqref{eq:jpol_R} with the aid of a systematic approach based on the Pauli-Villars regularization. This procedure is intrinsically free of potential spurious contributions and will be used to renormalize the remaining observable quantities.

\subsection{Current density via large-mass subtraction}

Before analyzing the current, we construct the following expansions of the QKE functions. We replace here momentum $\mathbf{p}$ and mass $m$ with $\lambda \mathbf{p}$ and $\lambda m$, respectively, and solve the QKEs~\eqref{eq:qke_f}--\eqref{eq:qke_vvec} as an asymptotic expansion in $1/\lambda$. The adiabatic energy takes the form
\begin{align}
\omega_\lambda (\mathbf{p}, t) &= \sqrt{\lambda^2 m^2 + [\lambda \mathbf{p} - e \mathbf{A}(t)]^2} \nonumber \\
{}&= \lambda p_0 + \mathcal{O} (\lambda^0).
\end{align}
The calculations are very similar to those in Sec.~\ref{sec:large_momentum}, and the result reads
\begin{widetext}
\begin{align}
f_\lambda (\mathbf{p}, t) & = \frac{e^2}{16\lambda^4 p_0^4} \, \bigg \{ \mathbf{E}^2(t) - \frac{[\mathbf{p} \mathbf{E}(t)]^2}{p_0^2} \bigg \} + \mathcal{O} \bigg ( \frac{1}{\lambda^5} \bigg ), \label{eq:large_lambda_f}\\
\mathbf{f}_\lambda (\mathbf{p}, t) & = \frac{e^2}{16\lambda^5 p_0^5} \, \bigg \{ \mathbf{E}(t) \times \dot{\mathbf{E}} (t) + \frac{[\mathbf{p} \dot{\mathbf{E}}(t)][\mathbf{p} \times \mathbf{E}(t)] - [\mathbf{p} \mathbf{E}(t)][\mathbf{p} \times \dot{\mathbf{E}}(t)]}{p_0 (p_0+m)} \bigg \} + \mathcal{O} \bigg ( \frac{1}{\lambda^6} \bigg ), \label{eq:large_lambda_f_vec} \\
\mathbf{u}_\lambda (\mathbf{p}, t) & = \frac{e}{8\lambda^3 p_0^3} \, \bigg \{ \dot{\mathbf{E}} (t) - \frac{[\mathbf{p} \dot{\mathbf{E}}(t)]\mathbf{p}}{p_0 (p_0+m)} \bigg \} + \mathcal{O} \bigg ( \frac{1}{\lambda^4} \bigg ), \label{eq:large_lambda_uvec} \\
\mathbf{v}_\lambda (\mathbf{p}, t) & = -\frac{e}{4\lambda^2 p_0^2} \, \bigg \{ \mathbf{E} (t) - \frac{[\mathbf{p} \mathbf{E}(t)]\mathbf{p}}{p_0 (p_0+m)} \bigg \} + \mathcal{O} \bigg ( \frac{1}{\lambda^3} \bigg ). \label{eq:large_lambda_vvec}
\end{align}
\end{widetext}
Note that Eqs.~\eqref{eq:large_p_f}--\eqref{eq:large_p_vvec} can be immediately obtained from Eqs.~\eqref{eq:large_lambda_f}--\eqref{eq:large_lambda_vvec} by setting $m = 0$ and $\lambda = 1$.

Having derived the above prerequisites, we will now inspect the unrenormalized current density~\eqref{eq:j_total_unrenorm}, which has the following form:
\begin{equation}
\mathbf{j} (t) = \int \! \frac{d\mathbf{p}}{(2\pi)^3} \, \boldsymbol{\mathcal{J}} (\mathbf{p}, m, t),
\label{eq:jJ}
\end{equation}
where we emphasize the explicit dependence on the fermion mass $m$. The integral~\eqref{eq:jJ} diverges logarithmically. According to the Pauli-Villars scheme, we introduce additional fermionic fields with large masses $M_i$. To regulate the logarithmic divergence, a single field of mass $M = \lambda m$ is sufficient. The regularized expression reads
\begin{equation}
\mathbf{j}^{(\text{PV})} (\lambda, t) = \int \! \frac{d\mathbf{p}}{(2\pi)^3} \, \big [ \boldsymbol{\mathcal{J}} (\mathbf{p}, m, t) - \boldsymbol{\mathcal{J}} (\mathbf{p}, \lambda m, t) \big ].
\label{eq:jJ_PV}
\end{equation}
For any given $\lambda$, this is finite, but the $\lambda \to \infty$ limit does not exist and requires renormalization. According to the conventional procedure of charge renormalization $e \to Z_3^{1/2} e$, we have to take into account the counterterm
\begin{equation}
j^{(\text{ct})}_\mu (\lambda, x) = \delta Z_3 (\lambda) \partial^\nu F_{\mu \nu} (x),
\end{equation}
which, in our setup, yields
\begin{equation}
\mathbf{j}^{(\text{ct})} (\lambda, t) = \delta Z_3 (\lambda) \dot{\mathbf{E}}(t).
\end{equation}
The renormalized current is then obtained via
\begin{equation}
\mathbf{j}_\text{R} (t) = \lim_{\lambda \to \infty} \big [ \mathbf{j}^{(\text{PV})} (\lambda, t) + \mathbf{j}^{(\text{ct})} (\lambda, t) \big ].
\label{eq:jR_PV-ct}
\end{equation}
First, we note that this construction is inherently gauge-invariant and cannot give rise to any spurious terms that do not vanish in the large-mass limit and differ in their form from the counterterm $\sim \dot{\mathbf{E}}(t)$. Second, the final result~\eqref{eq:jR_PV-ct} is governed by the choice of $\delta Z_3 (\lambda)$. Although it is clear that this function should cancel the divergent part of $\mathbf{j}^{(\text{PV})} (\lambda, t)$, its finite part should be determined by means of an appropriate physical renormalization condition. Let us introduce the following function:
\begin{multline}
\boldsymbol{\mathcal{J}}_\text{subtr} (\mathbf{p}, m, t) \equiv \lim_{\lambda \to \infty} \lambda^3 \boldsymbol{\mathcal{J}} (\lambda \mathbf{p}, \lambda m, t) \\
{} = \lim_{\lambda \to \infty} 4 e \lambda^3 \bigg \{ \mathbf{u}_\lambda (\mathbf{p},t) - \frac{[\mathbf{q}_\lambda \mathbf{u}_\lambda (\mathbf{p},t)]\mathbf{q}_\lambda}{\omega_\lambda (\omega_\lambda + \lambda m)} \bigg \},
\label{eq:Jsubtr_def}
\end{multline}
where $\mathbf{q}_\lambda \equiv \lambda \mathbf{p} - e \mathbf{A} (t)$. By using the expansion~\eqref{eq:large_lambda_uvec}, we find
\begin{equation}
\boldsymbol{\mathcal{J}}_\text{subtr} (\mathbf{p}, m, t) = \frac{e^2}{2p_0^3}  \, \bigg \{ \dot{\mathbf{E}} (t) - \frac{[\mathbf{p} \dot{\mathbf{E}}(t)]\mathbf{p}}{p_0^2} \bigg \}. \label{eq:j_subtr}
\end{equation}
Equation~\eqref{eq:jR_PV-ct} can be recast as
\begin{align}
\mathbf{j}_\text{R} (t) &= \int \! \frac{d\mathbf{p}}{(2\pi)^3} \, \big [ \boldsymbol{\mathcal{J}} (\mathbf{p}, m, t) - \boldsymbol{\mathcal{J}}_\text{subtr} (\mathbf{p}, m, t) \big ] \nonumber \\
{} &+ \lim_{\lambda \to \infty} \big [ \xi \ln \lambda + \delta Z_3 (\lambda) \big ] \dot{\mathbf{E}}(t).
\label{eq:jR_xi}
\end{align}
Here $\xi$ denotes the coefficient of the leading-order UV term in $\boldsymbol{\mathcal{J}} (\mathbf{p}, m, t)$ integrated over the angles:
\begin{equation}
\frac{1}{(2 \pi)^3} \int d\Omega_\mathbf{p} \, \boldsymbol{\mathcal{J}} (\mathbf{p}, m, t) = \frac{\xi \dot{\mathbf{E}}(t)}{|\mathbf{p}|^3} + \mathcal{O} \bigg ( \frac{1}{|\mathbf{p}|^4} \bigg ).
\end{equation}
The derivation of Eq.~\eqref{eq:jR_xi} is presented in Appendix~\ref{app_jR_xi}. The explicit form of $\xi$ is given by
\begin{equation}
\xi = \frac{2\alpha}{3\pi}.
\end{equation}
Therefore,
\begin{equation}
\delta Z_3 (\lambda) = -\frac{2\alpha}{3\pi} \ln \lambda + \text{const}.
\end{equation}
The remaining finite freedom is fixed by a physical renormalization condition. For our setup, the term $\dot{\mathbf{E}}(t)$ in the current density would correspond to an unphysical finite shift of the field-free vacuum polarization, violating the standard on-shell condition $\Pi_\text{R} (0) = 0$. To ensure the correct zero-field limit, one has to require that the term $\dot{\mathbf{E}}(t)$ be absent in Eq.~\eqref{eq:jR_xi}. This choice fixes
\begin{equation}
\delta Z_3 (\lambda) = -\frac{2\alpha}{3\pi} \ln \lambda
\label{eq:dZ3}
\end{equation}
and yields the compact prescription
\begin{equation}
\mathbf{j}_\text{R} (t) = \int \! \frac{d\mathbf{p}}{(2\pi)^3} \, \big [ \boldsymbol{\mathcal{J}} (\mathbf{p}, m, t) - \boldsymbol{\mathcal{J}}_\text{subtr} (\mathbf{p}, m, t) \big ].
\label{eq:jR_subtr_final}
\end{equation}
The renormalization constant~\eqref{eq:dZ3} reproduces the standard textbook result which provides a finite one-loop polarization function~\cite{blp}. The latter was used in the previous section to obtain the one-potential contribution, so the result~\eqref{eq:jR_subtr_final} agrees with the previous findings~\eqref{eq:jpol_R}--\eqref{eq:uR} given the explicit form of the subtraction term~\eqref{eq:j_subtr}. Finally, we note that if any spurious terms had needed to be subtracted, they would have been taken into account in our final prescription~\eqref{eq:jR_subtr_final}.

\subsection{Renormalized energy-momentum tensor}

Starting from the unrenormalized expression
\begin{equation}
T^{\mu \nu} (t) = \int \! \frac{d\mathbf{p}}{(2\pi)^3} \, \mathcal{T}^{\mu \nu} (\mathbf{p}, m, t),
\end{equation}
we first obtain the following (angle-averaged) subtraction terms:
\begin{align}
\mathcal{T}^{00}_\text{subtr} (\mathbf{p}, m, t) &\equiv \lim_{\lambda \to \infty} \lambda^3 \mathcal{T}^{00} (\lambda \mathbf{p}, \lambda m, t) \nonumber \\
{}&= \frac{e^2 \mathbf{E}^2(t)}{6 p_0^3} \, \bigg ( 1 + \frac{m^2}{2p_0^2} \bigg ), \label{eq:T00_lambda_subtr} \\
\mathcal{T}^{kl}_\text{subtr} (\mathbf{p}, m, t) &\equiv \lim_{\lambda \to \infty} \lambda^3 \mathcal{T}^{kl} (\lambda \mathbf{p}, \lambda m, t) \nonumber \\
{}&= - \frac{e^2}{3p_0^3} \bigg ( 1 - \frac{m^2}{p_0^2} \bigg ) \bigg [ E^k (t) E^l (t) \bigg ( 1 + \frac{m^2}{2p_0^2} \bigg ) \nonumber \\
{}&+ \frac{1}{2} \mathbf{E}^2 (t) g^{kl} \bigg ( 1 - \frac{m^2}{2p_0^2} \bigg ) \bigg ].
\end{align}
In the latter equation, we dropped gauge-dependent terms and odd contributions with respect to $\mathbf{p} \to -\mathbf{p}$ keeping in mind further integration over $\mathbf{p}$, where the gauge-dependent terms can be absorbed by the corresponding $\mathbf{p}$ shift in the integrals analogous to those discussed in Appendix~\ref{app_jR_xi}. The final subtraction prescriptions will only contain derivatives of $\mathbf{A}(t)$. The counterterms are given by
\begin{equation}
T^{(\text{ct})}_{\mu \nu} (\lambda, x) = \delta Z_3 (\lambda) \bigg [ F_{\mu \rho} (x) F_{\nu}{}^{\rho} (x)  - \frac{1}{4} g_{\mu \nu} F^2 (x) \bigg ],
\end{equation}
which yields
\begin{align}
T^{(\text{ct})}_{00} (\lambda, x) &= - \frac{1}{2} \delta Z_3 (\lambda) \mathbf{E}^2 (t), \label{eq:T00_ct} \\
T^{(\text{ct})}_{kl} (\lambda, x) &= \delta Z_3 (\lambda) \bigg [ E_k (t) E_l (t) + \frac{1}{2} \mathbf{E}^2 (t) g_{kl} \bigg ]. \label{eq:Tkl_ct}
\end{align}
Here we simply use the already established relation~\eqref{eq:dZ3}. The counterterms~\eqref{eq:T00_ct} and \eqref{eq:Tkl_ct} exactly cancel the divergent parts, and the renormalized energy density becomes
\begin{multline}
T^{00}_\text{R} (t) = \int \! \frac{d\mathbf{p}}{(2\pi)^3} \, \big [ \mathcal{T}^{00} (\mathbf{p}, m, t) - \mathcal{T}^{00}_\text{subtr} (\mathbf{p}, m, t) \big ] \\
{}= \int \! \frac{d\mathbf{p}}{(2\pi)^3} \, \bigg [ 4 \omega (\mathbf{p},t) f (\mathbf{p},t) - \frac{e^2 \mathbf{E}^2 (t)}{6 p_0^3} \bigg ( 1 + \frac{m^2}{2p_0^2} \bigg ) \bigg ].
\label{eq:T00R_subtr_final}
\end{multline}
Equivalently,
\begin{equation}
T^{00}_\text{R} (t) = 4 \int \! \frac{d\mathbf{p}}{(2\pi)^3} \, \omega (\mathbf{p},t)  f_\text{R} (\mathbf{p},t),  \label{eq:T00_R_explicit_sub}
\end{equation}
where $f_\text{R} = f - \delta f$ and
\begin{equation}
\delta f (\mathbf{p},t) \equiv \frac{e^2}{16 \omega^4} \, \bigg \{ \mathbf{E}^2(t) - \frac{[\mathbf{q} \mathbf{E}(t)]^2}{\omega^2} \bigg \}.
\end{equation}
One can straightforwardly show that the relation~\eqref{eq:energy_current} holds also for the renormalized quantities,
\begin{equation}
\dot{T}^{00}_\text{R} (t) = \mathbf{E} (t) \mathbf{j}_\text{R} (t).
\label{eq:energy_current_R}
\end{equation}
The renormalized stress tensor reads
\begin{align}
T^{kl}_\text{R} (t) &= 4 \int \! \frac{d\mathbf{p}}{(2\pi)^3} \, \bigg \{ \frac{q^k q^l}{\omega} \, \bigg [ f (\mathbf{p},t) - \frac{\mathbf{q} \mathbf{u} (\mathbf{p},t)}{\omega + m}\bigg ] \nonumber \\
{}&+ \frac{1}{2} \, \big [ q^k u^l (\mathbf{p},t) + q^l u^k (\mathbf{p},t) \big ] \nonumber \\
{}&+ \frac{e^2}{12p_0^3} \bigg ( 1 - \frac{m^2}{p_0^2} \bigg ) \bigg [ E^k (t) E^l (t) \bigg ( 1 + \frac{m^2}{2p_0^2} \bigg ) \nonumber \\
{}&+ \frac{1}{2} \mathbf{E}^2 (t) g^{kl} \bigg ( 1 - \frac{m^2}{2p_0^2} \bigg ) \bigg ] \bigg \}. \label{eq:Tkl_R_final}
\end{align}
It can also be written as
\begin{align}
T^{kl}_\text{R} (t) &= 4 \int \! \frac{d\mathbf{p}}{(2\pi)^3} \, \bigg \{ \frac{q^k q^l}{\omega} \, \bigg [ f_\text{R} (\mathbf{p},t) - \frac{\mathbf{q} \mathbf{u}_{\text{R}'} (\mathbf{p},t)}{\omega + m}\bigg ] \nonumber \\
{}&+ \frac{1}{2} \, \big [ q^k u^l_{\text{R}'} (\mathbf{p},t) + q^l u^k_{\text{R}'} (\mathbf{p},t) \big ] \bigg \}, \label{eq:Tkl_R_final_fR-uR}
\end{align}
where $\mathbf{u}_{\text{R}'} = \mathbf{u} - \delta \mathbf{u}'$ and
\begin{equation}
\delta \mathbf{u}' (\mathbf{p},t) = -\frac{3e^2}{8\omega^5} \bigg \{ \mathbf{E}(t) [\mathbf{q} \mathbf{E} (t)] - \frac{\mathbf{q} [\mathbf{q} \mathbf{E} (t)]^2}{\omega (\omega + m)} \bigg \}.
\label{eq:du2}
\end{equation}
This subtraction corresponds to the next-to-leading-order contribution to $\mathbf{u}_\lambda (\mathbf{p}, t)$. Note that it is odd in $\mathbf{q}$, whereas the leading-order term~\eqref{eq:uR} is even. The latter would not contribute to $T_\text{R}^{kl} (t)$, while the term~\eqref{eq:du2} would not contribute to $\mathbf{j}_\text{pol,R} (t)$ in Eq.~\eqref{eq:jpol_R_uR}. This means that in both observable quantities, one can safely use the following ``fully subtracted'' function:
\begin{equation}
\mathbf{u}_\text{R,full} (\mathbf{p},t) = \mathbf{u} (\mathbf{p},t) - \delta \mathbf{u} (\mathbf{p},t) - \delta \mathbf{u}' (\mathbf{p},t).
\label{eq:uR_full}
\end{equation}
Moreover, the subtracted function $f_{\mathrm R}(\mathbf{p},t)$ may also be used in the conduction current~\eqref{eq:jcond}, since $\delta f(\mathbf{p},t)$ is even in $\mathbf{q}$ and therefore gives no contribution upon momentum integration. The QKE formalism, supplemented with these subtraction prescriptions, provides a convenient and systematic framework for obtaining the physical observables. The charge renormalization procedure is carried out by means of the self-consistent counterterms and one universal constant $\delta Z_3$, which is chosen by one physical renormalization condition.

In what follows, we will briefly summarize the relevant results in the specific case of a linearly polarized external field, where the above expressions together with the QKE system itself take a simpler form. For this special case, subtraction prescriptions analogous to Eq.~\eqref{eq:uR_full} were obtained in Refs.~\cite{mamaev_trunov_1979,mostepanenko_1980,GMM}.

%%%%%%%%%%%%%%%%%%%%%%%%%%%%%%%%%%%%%%%%%%%%%%%%%%%%%%%%%

\section{Linear polarization: Summary} \label{sec:LP}

Here we specialize to an external electric field that remains parallel to a fixed unit vector $\mathbf{n}$ at all times. Accordingly, we write $\mathbf{E} (t) = E(t) \mathbf{n}$, where $E(t)$ is a scalar function.

As shown in Ref.~\cite{aleksandrov_kudlis_klochai}, it is convenient in this case to introduce a unit vector $\mathbf{e} (\mathbf{p}, t)$ by writing $\boldsymbol{\mu}_2 (\mathbf{p}, t) = \mu (\mathbf{p}, t) \mathbf{e} (\mathbf{p}, t)$. We then decompose the vector QKE functions into components parallel and perpendicular to $\mathbf{e} (\mathbf{p}, t)$: $\mathbf{f} = \tilde{f} \mathbf{e} + \mathbf{f}_\perp$, where $\tilde{f} = \mathbf{f} \mathbf{e}$, and similarly for $\mathbf{u}$ and $\mathbf{v}$. The fact that the external field is linearly polarized is equivalent to~\cite{aleksandrov_kudlis_klochai}
\begin{equation}
\mathbf{f} = \mathbf{0}~~\text{and}~~\mathbf{u}_\perp = \mathbf{v}_\perp = \mathbf{0}.
\label{eq:LP_statement}
\end{equation}
Under these conditions, the QKEs~\eqref{eq:qke_f}--\eqref{eq:qke_vvec} reduce to
\begin{align}
\dot{f} &= -2 \mu \tilde{u}, \label{eq:qke_LP_f} \\
\dot{\tilde{u}} &= \mu (2f - 1) + 2\omega \tilde{v}, \label{eq:qke_LP_u} \\
\dot{\tilde{v}} &= -2\omega \tilde{u}, \label{eq:qke_LP_v}
\end{align}
where
\begin{equation}
\mu = \mu (\mathbf{p}, t) = \frac{eE(t) \pi_\perp (\mathbf{p})}{2\omega^2 (\mathbf{p} , t)} \,.
\end{equation}
Here $\pi_\perp (\mathbf{p}) \equiv \sqrt{m^2 + \mathbf{p}_\perp^2}$, and $\mathbf{p}_\perp = \mathbf{q}_\perp$ is the transverse momentum component, i.e. that perpendicular to $\mathbf{n}$. The longitudinal momentum is $q_\parallel = \mathbf{q} \mathbf{n} = p_\parallel - e A(t)$ (not to be confused with $\mathbf{q} \mathbf{e} = -mq_\parallel/\pi_\perp$). The quasienergy satisfies $\omega^2 = q_\parallel^2 + \pi_\perp^2$. The QKE functions are independent of the azimuthal angle of $\mathbf{p}_\perp$. The system~\eqref{eq:qke_LP_f}--\eqref{eq:qke_LP_v} has been discussed and numerically implemented in numerous studies (see, e.g., Refs.~\cite{sevostyanov_prd_2021, schmidt_1998, kluger_prd_1998, schmidt_prd_1999, bloch_prd_1999, blaschke_prd_2011, aleksandrov_epjst_2020, alkofer_prl_2001, otto_plb_2015, panferov_epjd_2016, aleksandrov_symmetry, aleksandrov_sevostyanov_2025,brass_arxiv_2025}). We also note that it is often formulated in terms of $f' = f$, $u' = -2\tilde{u}$, and $v' = 2 \tilde{v}$. The integral of motion~\eqref{eq:qke_int_of_motion_gen} takes the following form:
\begin{equation}
\frac{1}{4} (1-2f)^2 + \tilde{u}^2 + \tilde{v}^2 = \frac{1}{4}.
\end{equation}

The remaining steps are straightforward algebra. The subtraction terms for $\mathbf{u}$ are also parallel to $\mathbf{e}$, so we define
\begin{align}
f_\text{R} (\mathbf{p}, t) &= f(\mathbf{p}, t) - \frac{e^2 E^2(t) \pi^2_\perp (\mathbf{p})}{16 \omega^6 (\mathbf{p}, t)},\\
\tilde{u}_\text{R} (\mathbf{p}, t) &= \tilde{u} (\mathbf{p}, t) + \frac{e\dot{E}(t) \pi_\perp (\mathbf{p})}{8 \omega^4 (\mathbf{p}, t)} \nonumber \\
{}& - \frac{3e^2 E^2(t) q_\parallel \pi_\perp (\mathbf{p})}{8\omega^6 (\mathbf{p}, t)}. \label{eq:LP_uR}
\end{align}
The last two terms are even and odd, respectively. Note that Eq.~\eqref{eq:LP_uR} is an analog of the function~\eqref{eq:uR_full}, but we use a shorter subscript ``R'' instead of ``R,\,full''. In a compact form, one can write
\begin{align}
f_\text{R} (\mathbf{p}, t) &= f(\mathbf{p}, t) - \frac{\mu^2 (\mathbf{p}, t)}{4\omega^2 (\mathbf{p}, t)},\\
\tilde{u}_\text{R} (\mathbf{p}, t) &= \tilde{u} (\mathbf{p}, t) + \frac{1}{4\omega (\mathbf{p}, t)} \frac{d}{dt} \frac{\mu (\mathbf{p}, t)}{\omega (\mathbf{p}, t)}.
\end{align}
Note that $\dot{f}_\text{R} = -2\mu \tilde{u}_\text{R}$. The conduction current is given by
\begin{equation}
\mathbf{j}_\text{cond} (t) = 4e \int \! \frac{d\mathbf{p}}{(2\pi)^3} \, \frac{\mathbf{q}}{\omega} \, f (\mathbf{p},t),
\end{equation}
where $f$ can be replaced with $f_\text{R}$ as the subtraction term, being even in $\mathbf{q}$, does not contribute. The transverse current vanishes by symmetry: the integrand is odd under $\mathbf{p}_\perp \to -\mathbf{p}_\perp$. Therefore,
\begin{equation}
\mathbf{j}_\text{cond} (t) = 4e \mathbf{n} \int \! \frac{d\mathbf{p}}{(2\pi)^3} \, \frac{q_\parallel}{\omega} \, f (\mathbf{p},t),
\end{equation}
The integrals over $\mathbf{p}$ are practically two-dimensional: $d\mathbf{p} = 2\pi |\mathbf{p}_\perp|d|\mathbf{p}_\perp|dp_\parallel$. The renormalized polarization current has the form
\begin{equation}
\mathbf{j}_\text{pol,R} (t) = 4e \int \! \frac{d\mathbf{p}}{(2\pi)^3} \frac{q_\parallel \mathbf{p}_\perp - \pi^2_\perp (\mathbf{p}) \mathbf{n}}{\pi_\perp (\mathbf{p}) \omega (\mathbf{p}, t)} \, \tilde{u}_\text{R} (\mathbf{p}, t).
\end{equation}
The first term with $\mathbf{p}_\perp$ is odd with respect to $\mathbf{p}_\perp \to -\mathbf{p}_\perp$, so it can be discarded:
\begin{equation}
\mathbf{j}_\text{pol,R} (t) = -4e \mathbf{n} \int \! \frac{d\mathbf{p}}{(2\pi)^3} \frac{\pi_\perp (\mathbf{p})}{\omega (\mathbf{p}, t)} \, \tilde{u}_\text{R} (\mathbf{p}, t).
\label{eq:jpol_R_LP}
\end{equation}
Thus, the induced current density is always collinear with the external field $\mathbf{E}\parallel \mathbf{n}$.

The energy density has the following expression:
\begin{equation}
T^{00}_\text{R} (t) = 4 \int \! \frac{d\mathbf{p}}{(2\pi)^3} \, \omega (\mathbf{p},t)  f_\text{R} (\mathbf{p},t).
\end{equation}
Defining a scalar function $j_\text{R}(t)$ by
\begin{equation}
\mathbf{j}_\text{cond} (t) + \mathbf{j}_\text{pol,R} (t) = j_\text{R} (t) \mathbf{n},
\end{equation}
one readily verifies the energy-balance relation
\begin{equation}
\dot{T}^{00}_\text{R} (t) = E(t) j_\text{R} (t).
\end{equation}
The renormalized stress tensor is given by
\begin{align}
T^{kl}_\text{R} (t) &= 4 \int \! \frac{d\mathbf{p}}{(2\pi)^3} \, \bigg \{ \frac{q^k q^l}{\omega} \, \bigg [ f_\text{R} (\mathbf{p},t) + \frac{mq_\parallel \tilde{u}_\text{R} (\mathbf{p}, t)}{\pi_\perp (\omega + m)} \bigg ] \nonumber \\
{}&+ \frac{1}{2} \, \big [ q^k e^l (\mathbf{p},t) + q^l e^k (\mathbf{p},t) \big ] \tilde{u}_\text{R} (\mathbf{p}, t) \bigg \}.
\end{align}
This tensor can be represented in a more transparent form if we specify the Cartesian projections of the field polarization vector $\mathbf{n}$. Assuming that $\mathbf{n}$ points at the $z$ direction, we arrive at
\begin{align}
T^{11}_\text{R} (t) &= 2 \int \! \frac{d\mathbf{p}}{(2\pi)^3} \, \frac{\mathbf{p}^2_\perp}{\omega} \bigg [ f_\text{R} (\mathbf{p},t) + \frac{q_\parallel}{\pi_\perp} \tilde{u}_\text{R} (\mathbf{p}, t) \bigg ], \label{eq:T11_LP} \\
T^{22}_\text{R} (t) &= T^{11}_\text{R} (t), \\
T^{33}_\text{R} (t) &= 4 \int \! \frac{d\mathbf{p}}{(2\pi)^3} \, \frac{q_\parallel}{\omega} \bigg [ q_\parallel f_\text{R} (\mathbf{p},t) - \pi_\perp \tilde{u}_\text{R} (\mathbf{p}, t) \bigg ].
\label{eq:T33_LP}
\end{align}
All remaining components vanish. We note a sign difference relative to Ref.~\cite{GMM} in the terms proportional to $\tilde{u}_\text{R}$ in the expressions analogous to Eqs.~\eqref{eq:jpol_R_LP}, \eqref{eq:T11_LP}--\eqref{eq:T33_LP}.

In the case of a linearly polarized field, the spin-resolved distributions of particles are completely symmetric, and the net spin density vanishes, $\mathbf{S} (t) = 0$.

%%%%%%%%%%%%%%%%%%%%%%%%%%%%%%%%%%%%%%%%%%%%%%%%%%%%%%%%%

\section{Conclusion} \label{sec:conclusion}

In this paper, we have presented a detailed study of vacuum polarization and electron-positron pair production in a spatially homogeneous, time-dependent electric field of arbitrary polarization within the quantum-kinetic-equation (QKE) framework. Building on the adiabatic-basis formulation developed in Ref.~\cite{aleksandrov_kudlis_klochai}, we analyzed the structure and properties of the resulting closed system of kinetic equations. In addition to momentum-resolved particle distributions, we considered observable expectation values induced by the background field. In particular, we expressed the electric current density, the energy-momentum tensor, and the angular-momentum (spin) tensor in terms of the QKE functions and clarified their correspondence to the Dirac-Heisenberg-Wigner (DHW) description.

The logarithmic ultraviolet divergence appearing in the polarization current was removed using two complementary approaches: renormalization of the one-potential contribution and large-mass subtraction within the Pauli-Villars scheme. We described both procedures in detail and verified their mutual consistency. This analysis also enabled us to derive finite, gauge-invariant expressions for the energy-momentum tensor.

For practical applications, we also summarized the reduced kinetic system and collected compact formulas for the relevant observables in the case of linear polarization. Overall, the framework developed here provides an efficient and transparent tool for quantitative studies of nonperturbative QED dynamics in strong, time-dependent electric fields.

%%%%%%%%%%%%%%%%%%%%%%%%%%%%%%%%%%%%%%%%%%%%%%%%%%%%%%%%%

\acknowledgments

We are grateful to Prof.~V.~M.~Shabaev for valuable discussions. The study was funded by the Foundation for the Advancement of Theoretical Physics and Mathematics BASIS (Project No.~25-1-3-48-1). The work of A.K. was supported by the Icelandic Research Fund (Ranns\'oknasj\'o{\dh}ur, Grant No.~2410550).

%%%%%%%%%%%%%%%%%%%%%%%%%%%%%%%%%%%%%%%%%%%%%%%%%%%%%%%%%

\appendix

\section{Connection with the DHW functions in Refs.~\cite{aleksandrov_kudlis_klochai,BB_prd_1991,hebenstreit_prd_2010}} \label{app:BB}

For convenience, we list the explicit relations between the QKE functions $f$, $\mathbf{f}$, $\mathbf{u}$, and $\mathbf{v}$ and the functions involved in the DHW formalism~\cite{BB_prd_1991,hebenstreit_prd_2010}. These relations were derived in Ref.~\cite{aleksandrov_kudlis_klochai}. They are
\begin{align}
\tilde{\mathfrak{s}} - \tilde{\mathfrak{s}}_{\mathbf{A}=\mathbf{0}} &= 4 \bigg ( \frac{m}{\omega} f - \frac{\mathbf{q} \mathbf{u}}{\omega} \bigg ) , \label{eq:rel_s}\\
\tilde{\boldsymbol{\mathfrak{v}}} - \tilde{\boldsymbol{\mathfrak{v}}}_{\mathbf{A}=\mathbf{0}} &= 4 \bigg [ \frac{\mathbf{q}}{\omega} f + \mathbf{u} - \frac{\mathbf{q} (\mathbf{q} \mathbf{u})}{\omega (\omega+m)} \bigg ] , \label{eq:rel_v}\\
\tilde{\boldsymbol{\mathfrak{a}}} &= -4 \bigg [ \frac{m}{\omega} \mathbf{f} - \frac{\mathbf{q} \times \mathbf{v}}{\omega} + \frac{\mathbf{q} (\mathbf{q} \mathbf{f})}{\omega(\omega+m)} \bigg ] , \label{eq:rel_a} \\
\tilde{\boldsymbol{\mathfrak{t}}} &= -4 \bigg [ \frac{m}{\omega} \mathbf{v} - \frac{\mathbf{q} \times \mathbf{f}}{\omega} + \frac{\mathbf{q} (\mathbf{q} \mathbf{v})}{\omega (\omega + m)} \bigg ] . \label{eq:rel_t1} 
\end{align}
The DHW functions on the left-hand sides follow the notation of Refs.~\cite{aleksandrov_kudlis_klochai,hebenstreit_prd_2010}. Reference~\cite{BB_prd_1991} uses a different notation, namely: $f_3 = \tilde{\mathfrak{s}}$, $\mathbf{g}_0 = -\tilde{\boldsymbol{\mathfrak{a}}}$, $\mathbf{g}_1 = \tilde{\boldsymbol{\mathfrak{v}}}$, and $\mathbf{g}_2 = \tilde{\boldsymbol{\mathfrak{t}}}$.

Unrenormalized expressions for several observables were obtained in Refs.~\cite{BB_prd_1991,hebenstreit_prd_2010}. In particular, the current density, energy density, and spin density in Eqs.~(33), (37), and (41) of Ref.~\cite{BB_prd_1991} agree with our results in Sec.~\ref{sec:unrenorm}. For instance, the total current density, i.e. the sum of Eqs.~\eqref{eq:jcond} and \eqref{eq:jpol}, coincides up to the factor $e/(2\pi)^3$ with the $\mathbf{p}$ integral of Eq.~\eqref{eq:rel_v}, which is consistent with Eq.~(33) of Ref.~\cite{BB_prd_1991} and supports the interpretation of $\tilde{\boldsymbol{\mathfrak{v}}}$ ($\mathbf{g}_1$ in Ref.~\cite{BB_prd_1991}) as the current phase-space density.

%%%%%%%%%%%%%%%%%%%%%%%%%%%%%%%%%%%%%%%%%%%%%%%%%%%%%%%%%

\section{Direct calculation of the one-potential contribution to the current density} \label{app_1P_direct}

Here we treat the external electric field perturbatively and work in the interaction picture, where the evolution operator has the form
\begin{equation}
U(t, t_0) = \mathcal{T} \, \mathrm{exp} \Bigg [ \!\! -i \int \limits_{t_0}^{t} \! dx^0 \! \int \! d\mathbf{x} \, j^\nu_0 (x) A_\nu (x) \Bigg ]. \label{eq:S_interaction_picture}
\end{equation}
Here $\mathcal{T}$ denotes time ordering, $A_\nu (x) = A_\nu (x^0)$ is the external vector potential, and $j^\nu_0 (x)$ is the current operator of the {\it free} Dirac field. The vacuum expectation value of the current density~\eqref{eq:j_mean} is then
\begin{equation}
j^\mu (x^0) = \frac{1}{V} \int \! d\mathbf{x} \, \langle 0 | U^\dagger (x^0, t_\text{in}) j^\mu_0 (x) U (x^0, t_\text{in}) | 0 \rangle.
\end{equation}
The first-order (one-potential) contribution reads
\begin{align}
j^{(\text{1P})}_{\mu} (x^0) &= \frac{2}{V} \, \mathrm{Im} \int \! d\mathbf{x} \! \int \limits_{t_\text{in}}^{x^0} \! dy^0 \! \int \! d\mathbf{y}  A_\nu (y^0) \langle 0 | j_{0\mu} (x) j^\nu_0 (y) | 0 \rangle \nonumber \\
{}&= \frac{2 e^2}{V} \, \mathrm{Im} \int \! d\mathbf{x} \! \int \limits_{t_\text{in}}^{x^0} \! dy^0 \! \int \! d\mathbf{y}  A_\nu (y^0) \nonumber \\
{}&\times \mathrm{Tr} \big [ \gamma_\mu S_0 (x,y) \gamma^\nu S_0 (y,x) \big ].
\label{eq:app_j_1P}
\end{align}
Here $S_0 (x,y)$ is the free Dirac propagator:
\begin{equation}
S_0(x,y) = \int \! \frac{d\mathbf{p}}{(2\pi)^3} \, \mathrm{e}^{i \mathbf{p} (\mathbf{x} - \mathbf{y})} i S_0(x^0, y^0, \mathbf{p}),
\end{equation}
where
\begin{multline}
S_0(x^0, y^0, \mathbf{p}) = \theta (x^0-y^0) \, \frac{\gamma^0 p_0 - \boldsymbol{\gamma} \mathbf{p} + m}{2p^0} \, \mathrm{e}^{-i p^0 (x^0 - y^0)} \\
{}- \theta (y^0-x^0) \, \frac{\gamma^0p_0 + \boldsymbol{\gamma} \mathbf{p} - m}{2p^0} \, \mathrm{e}^{ip^0 (x^0 - y^0)}
\end{multline}
and $p^0 = \sqrt{m^2 + \mathbf{p}^2}$. By integrating over $\mathbf{y}$ and $\mathbf{x}$, we find
\begin{multline}
j^{(\text{1P})}_{\mu} (t) = 2e^2 \, \mathrm{Im} \int \limits_{-\infty}^{t} \! dt' \! \int \! \frac{d\mathbf{p}}{(2\pi)^3} \, \mathbf{A} (t') \\
{}\times \mathrm{Tr} \, \big [ \gamma_\mu S_0 (t, t', \mathbf{p}) \boldsymbol{\gamma} S_0(t', t, \mathbf{p}) \big ].
\end{multline}
First, the trace vanishes for $\mu=0$; hence $j^{(\text{1P})}_0 (t) = 0$. Second, for $\mu = k \in \{1, 2, 3\}$, the integrand contains the factor $\mathrm{exp} [-2ip^0 (t-t')]/(4p_0^2)$ and the following trace:
\begin{multline}
\mathrm{Tr} \, \big [ \gamma^k (p^0\gamma^0 - \boldsymbol{\gamma} \mathbf{p} + m) \gamma^j (-p^0\gamma^0 - \boldsymbol{\gamma} \mathbf{p} + m) \big ] \\
{}= 8[p^k p^j - p_0^2 \delta^{kj}].
\end{multline}
Integrating by parts with respect to $t'$ yields Eq.~\eqref{eq:jpol_1P}.

%%%%%%%%%%%%%%%%%%%%%%%%%%%%%%%%%%%%%%%%%%%%%%%%%%%%%%%%%

\section{Derivation of Eq.~\eqref{eq:jR_xi}} \label{app_jR_xi}

According to Eq.~\eqref{eq:jR_PV-ct}, the renormalized current reads
\begin{multline}
\mathbf{j}_\text{R} (t) = \lim_{\lambda \to \infty} \bigg \{ \int \! \frac{d\mathbf{p}}{(2\pi)^3} \, \big [ \boldsymbol{\mathcal{J}} (\mathbf{p}, m, t) - \boldsymbol{\mathcal{J}} (\mathbf{p}, \lambda m, t) \big ]\\
{}+\delta Z_3 (\lambda) \dot{\mathbf{E}}(t) \bigg \}.
\label{eq:app_jR1}
\end{multline}
For fixed $\lambda$, the momentum integral can be decomposed into three finite terms:
\begin{align}
&\int \! \frac{d\mathbf{p}}{(2\pi)^3} \, \big [ \boldsymbol{\mathcal{J}} (\mathbf{p}, m, t) - \boldsymbol{\mathcal{J}}_\text{subtr} (\mathbf{p}, m, t) \big ] \nonumber \\
{}&+\int \! \frac{d\mathbf{p}}{(2\pi)^3} \, \big [ \boldsymbol{\mathcal{J}}_\text{subtr} (\mathbf{p}, m, t) - \lambda^3 \boldsymbol{\mathcal{J}} (\lambda \mathbf{p}, \lambda m, t) \big ] \nonumber \\
{}&+\int \! \frac{d\mathbf{p}}{(2\pi)^3} \, \big [ \lambda^3 \boldsymbol{\mathcal{J}} (\lambda \mathbf{p}, \lambda m, t) - \boldsymbol{\mathcal{J}} (\mathbf{p}, \lambda m, t) \big ],
\end{align}
where $\boldsymbol{\mathcal{J}}_\text{subtr} (\mathbf{p}, m, t)$ is defined in Eq.~\eqref{eq:Jsubtr_def}. The first term is $\lambda$-independent, the second vanishes as $\lambda \to \infty$, and the third can be evaluated as
\begin{align}
&\lim_{\Lambda \to \infty} \int\limits_0^{\Lambda} \! \frac{d|\mathbf{p}| |\mathbf{p}|^2}{(2\pi)^3} \int d\Omega_\mathbf{p} \, \big [ \lambda^3 \boldsymbol{\mathcal{J}} (\lambda \mathbf{p}, \lambda m, t) - \boldsymbol{\mathcal{J}} (\mathbf{p}, \lambda m, t) \big ] \nonumber \\
{}&= \lim_{\Lambda \to \infty} \int\limits_{\Lambda}^{\lambda \Lambda} \! \frac{d|\mathbf{p}| |\mathbf{p}|^2}{(2\pi)^3} \int d\Omega_\mathbf{p} \, \boldsymbol{\mathcal{J}} (\mathbf{p}, \lambda m, t) \nonumber \\
{}&= \lim_{\Lambda \to \infty} \int\limits_{\Lambda}^{\lambda \Lambda} \! d|\mathbf{p}| \bigg [ \frac{\xi \dot{\mathbf{E}}(t)}{|\mathbf{p}|} + \mathcal{O} \bigg ( \frac{1}{|\mathbf{p}|^2} \bigg ) \bigg ] = \xi \dot{\mathbf{E}}(t) \ln \lambda.
\label{eq:app_Lambda}
\end{align}
By combining Eqs.~\eqref{eq:app_jR1}--\eqref{eq:app_Lambda}, one obtains Eq.~\eqref{eq:jR_xi}.

%%%%%%%%%%%%%%%%%%%%%%%%%%%%%%%%%%%%%%%%%%%%%%%%%%%%%%%%%

\end{document}